\documentclass[pra,aps,amsfonts,groupedaddress,notitlepage,nofootinbib,longbibliography,twocolumn,10pt]{revtex4-2}
\usepackage[utf8]{inputenc}
\usepackage[dvipsnames]{xcolor}
\usepackage[inline]{enumitem}
\usepackage{anyfontsize}
\usepackage{amsmath,amssymb,amsfonts,mathtools,dsfont,relsize}
\usepackage{microtype} 
\usepackage{graphicx}

\usepackage{amsthm}

\usepackage{soul} 

\usepackage[dvipsnames]{xcolor}
\usepackage{stmaryrd} 
\usepackage{suffix} 
\usepackage{comment}
\usepackage{xpatch}
\makeatletter
\patchcmd{\@ssect@ltx}
    {\addcontentsline{toc}{#1}{\protect\numberline{}#8}}
    {}
    {}
    {}
\makeatother

\renewcommand{\thesection}{\arabic{section}}
\renewcommand{\thesubsection}{\thesection.\arabic{subsection}}
\renewcommand{\thesubsubsection}{\thesubsection.\arabic{subsubsection}}
\makeatletter
\renewcommand{\p@subsection}{}
\def\l@subsubsection#1#2{}
\makeatother

\usepackage{adjustbox}

\makeatletter
\g@addto@macro\bfseries{\boldmath}
\makeatother

\newcommand{\vpd}[0]{\vphantom{\dagger}}
\newcommand{\vps}[0]{\vphantom{*}}

\DeclarePairedDelimiter{\abs}{\lvert}{\rvert}
\DeclarePairedDelimiter{\expval}{\langle}{\rangle}

\DeclarePairedDelimiter{\ket}{\lvert}{\rangle}
\DeclarePairedDelimiter{\bra}{\langle}{\rvert}

\newcommand{\ii}[0]{\mathrm{i}}
\newcommand{\ee}[0]{\mathrm{e}}
\newcommand{\bvec}[1]{\boldsymbol{#1}}

\newcommand{\thed}[0]{\mathrm{d}}

\DeclareMathOperator{\Wg}{Wg}
\DeclareMathOperator{\var}{var}
\DeclareMathOperator{\Avg}{avg}

\newcommand{\kron}[1]{\delta_{#1}}

\newcommand{\ident}[0]{\mathds{1}}

\newcommand{\Reals}{\mathbb{R}}
\newcommand{\Comps}{\mathbb{C}}

\newcommand{\Nats}{\mathbb{N}}

\newcommand{\Orth}[1]{\mathsf{O} ( #1 )}

\newcommand{\Unitary}[1]{\mathsf{U} ( #1 )}
\newcommand{\Symp}[1]{\mathsf{Sp} ( #1 )}

\newcommand{\Order}[1]{O ( #1 )}

\catcode`,\active

\catcode`\,12

\catcode`,\active

\catcode`\,12

\newcommand{\BKop}[2]{\ket{#1} \hspace{-0.2mm} \bra{#2}}

\newcommand{\inprod}[2]{ \left\langle #1 \middle| #2 \right\rangle}
\newcommand{\matel}[3]{\left\langle #1 \middle| #2 \middle| #3 \right\rangle}
\WithSuffix\newcommand\matel*[3]{\langle #1 | #2 | #3 \rangle}

\DeclareMathOperator*{\tr}{tr}
\DeclareMathOperator*{\trace}{tr}

\newcommand{\Hilbert}[0]{\mathcal{H}}

\DeclareMathOperator{\expec}{\mathbb{E}}

\usepackage{newtxtext}
\usepackage[smallerops]{newtxmath}

\DeclareMathAlphabet{\mathcal}{OMS}{cmsy}{m}{n}
\DeclareMathAlphabet{\mathsf}{OT1}{cmss}{m}{n}

\SetSymbolFont{stmry}{bold}{U}{stmry}{m}{n}

\usepackage{silence}
\usepackage[colorlinks={true}, 
linkcolor={BrickRed},     
citecolor={MidnightBlue}, 
urlcolor={PineGreen},     
bookmarks={true},
pdftitle={GUE noise},
pdffitwindow={true},
pdfpagetransition = {Fade},
pdfcenterwindow={true}
]{hyperref} 

\DeclareSymbolFont{largesymbolstx}{OMX}{txex}{m}{n}
\DeclareMathSymbol{\txintop}{\mathop}{largesymbolstx}{"52}
\DeclareRobustCommand\int{\txintop\nolimits}

\renewcommand{\lambda}{\lambdaup}

\makeatletter
\g@addto@macro\bfseries{\boldmath}
\makeatother

\begin{document}

\title{Exact analytic toolbox for quantum dynamics with tunable noise strength}

\author{Mert Okyay\,\href{https://orcid.org/0000-0002-5883-4919}{\includegraphics[width=6.5pt]{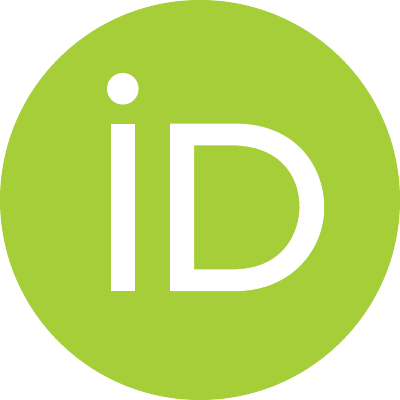}}}
\email{mert.okyay@colorado.edu}
\author{Oliver Hart\,\href{https://orcid.org/0000-0002-5391-7483}{\includegraphics[width=6.5pt]{imgs/ORCIDiD_iconvector.pdf}}}
\author{Rahul Nandkishore\,\href{https://orcid.org/0000-0001-5703-6758}{\includegraphics[width=6.5pt]{imgs/ORCIDiD_iconvector.pdf}}}
\author{Aaron J. Friedman\,\href{https://orcid.org/0000-0003-2159-6475}{\includegraphics[width=6.5pt]{imgs/ORCIDiD_iconvector.pdf}}}
\affiliation{Department of Physics and Center for Theory of Quantum Matter, University of Colorado Boulder, Boulder, Colorado 80309 USA}

\begin{abstract}
We introduce a framework that allows for the exact analytic treatment of quantum dynamics subject to coherent noise. The noise is modeled via unitary evolution under a Hamiltonian drawn from a random-matrix ensemble for arbitrary Hilbert-space dimension $N$.  While the methods we develop apply to generic such ensembles with a notion of rotation invariance, we focus largely on the Gaussian unitary ensemble (GUE). Averaging over the ensemble of ``noisy'' Hamiltonians produces an effective quantum channel, the properties of which are analytically calculable and determined by the spectral form factors of the relevant ensemble. We compute spectral form factors of the GUE exactly for any finite $N$, along with the corresponding GUE quantum channel, and its variance. Key advantages of our approach include the ability to access exact analytic results for any $N$ and the ability to tune to the noise-free limit (in contrast, e.g., to the Haar ensemble), and analytic access to moments beyond the variance. We also highlight some unusual features of the GUE channel, including the \emph{nonmonotonicity} of the coefficients of various operators as a function of noise strength and the failure to saturate the Haar-random limit, even with infinite noise strength. 
\end{abstract}

\maketitle


\section{Introduction}
\label{sec:intro}

The advent of quantum hardware with experimentally controllable qubits has led to 
significant interest in quantum dynamics. Of particular interest is how these synthetic quantum systems can be used to realize new capabilities for near-term technologies. These efforts include identifying quantum algorithms that outperform classical analogues~\cite{PreskillSupreme, Preskill2018quantumcomputing,  QC_book}, 
quantum error correction (QEC) for the storage of logical qubits~\cite{CalderbankGood, SteaneQEC1, SteaneQEC2, DiVincenzoShorQEC, Kitaev97, GottesmanFTQC, Knill_2000, Dennis_2002, TemmeZNE, Lieu2020}, the preparation of many-body resource states~\cite{QC_book, Cruz_2019, Satzinger_2021, SpeedLimit, Lu_2022}, controlled simulations of quantum dynamics 
of real materials~\cite{HaahSim2021}, learning about unknown quantum states and processes~\cite{HuangShadow, HuangAdvantage, HuangEfficient}, developing precision sensors for, e.g., metrology~\cite{metro_rev, squeeze_rev}, and even playing nonlocal quantum games~\cite{DanielStringOrder, BulchandaniGames, BBSgame, Daniel2022Exp, hart2024playing}.

While quantum technologies can realize significant advantages over their classical analogues (e.g., quantum computation), they are generally far more sensitive to sources of error~\cite{PreskillSupreme, Preskill2018quantumcomputing,  QC_book}. Accordingly, a crucial area of research is the development of strategies for mitigating and correcting for such errors. In the context of stabilizer QEC~\cite{CalderbankGood, SteaneQEC1, SteaneQEC2, DiVincenzoShorQEC, Kitaev97, GottesmanFTQC, Knill_2000, Dennis_2002, TemmeZNE, Lieu2020}, e.g., it is often the case that \emph{arbitrary} local errors can be corrected via measurements and outcome-dependent feedback \emph{before} they can proliferate and produce a logical error~\cite{QC_book, AQM}. Beyond this setting, the effects of noise on quantum systems---and strategies for their mitigation---must be considered on a case-by-case basis.

However, outside of the context of stabilizer QEC~\cite{CalderbankGood, SteaneQEC1, SteaneQEC2, DiVincenzoShorQEC, Kitaev97, GottesmanFTQC, Knill_2000, Dennis_2002, TemmeZNE, Lieu2020}, 
tools for the analytical treatment of noisy quantum dynamics are generally lacking. While the Lindblad description~\cite{lindblad1973entropy, Lindblad, QC_book, OpenSysBook, DecoherenceBook, AQM} of open quantum systems describes decohering effects due to interaction of a quantum system with a Markovian  environment, the \emph{coherent} sources of error of interest herein are generally modeled using noisy numerical simulations. Accordingly, we expect the framework we present herein to be useful to the description of dynamics on the noisy intermediate-scale quantum (NISQ) devices presently available.

In this work, we present an analytical toolbox for describing quantum dynamics in the presence of coherent sources of noise with tunable strength. The noisy channel may act on a system of interest, subsets thereof, or the combination of the system and ``environmental'' degrees of freedom; symmetries and/or constraints may be included using projectors~\cite{U1FRUC, ConstrainedRUC, MIPT}. Exact results are furnished by drawing the unitary evolution operator from ensembles common in random matrix theory (RMT)~\cite{guhr1998random, mehta2004random, TaoRMT, Brezin:2016eax, YoshidaCBD, eynard2018random, Livan_2018}. In particular, our results apply to noisy evolution generated by random $N \times N$ Hamiltonians $G$ drawn from an ensemble $\mathcal{E}$ whose probability distribution function (PDF) $P_{\mathcal{E}} (G)$ is rotation invariant---i.e., $P_{\mathcal{E}}(U G U^\dagger) = P_{\mathcal{E}}(G)$ for all $U$ belonging to some matrix group $\mathsf{R}$. The noise strength is captured either by the duration $t$ of time evolution generated by $G$ or, equivalently, the standard deviation $\sigma$ of the distribution $P_{\mathcal{E}} (G)$. 

For concreteness, we explicitly consider Hamiltonians $G$ drawn from the Gaussian unitary ensemble (GUE)~\cite{mehta2004random, Livan_2018, eynard2018random, YoshidaCBD, TaoRMT, Brezin:2016eax}. We derive an expression for the average evolution of any operator $A$ or density matrix in Sec.~\ref{sec:channel def}; the result contains two $t$-dependent terms, proportional to $A$ and $\tr (A) \ident$, respectively, effectively realizing a \emph{depolarizing} channel~\cite{QC_book} whose strength depends only on $\smash{\sigma^2 t^2}$, where $\smash{\sigma^2}$ is the variance of matrix elements of $G$. This time dependence is related to the two-point spectral form factor (SFF) $\mathcal{R}_2 (t)$~\cite{RMT_SFF, YoshidaSFF, ShenkerRMT, CDLC1,  U1FRUC, ConstrainedRUC, MIPT}, a standard diagnostic of spectral rigidity (i.e., level repulsion) in quantum systems, and a common diagnostic in the field of quantum chaos~\cite{RMT_SFF, YoshidaSFF, ShenkerRMT, CDLC1,  U1FRUC, ConstrainedRUC, MIPT, Micklitz}. We compute the two-point SFF for the GUE \emph{exactly} for any finite $N$, correcting typos that appear in Ref.~\citenum{delCampoGUESFF}.
We also compute the variance of the noisy channel, which can be used to diagnose, e.g., typicality of the average. We also describe the evaluation of arbitrary cumulants, where the $k$th cumulant depends on the $2k$-point and lower SFFs; in Sec.~\ref{sec:SFF} we compute exactly the first few SFFs for the GUE for arbitrary Hilbert-space dimension $N$, and also detail how to compute all SFFs. We expect these results to be relevant to quantum chaos \cite{BohigasChaos, RMT_SFF, ShenkerRMT, YoshidaSFF, CDLC1, U1FRUC, ConstrainedRUC, MIPT, AmitQuantErgo} and even the study of black holes \cite{ShenkerRMT, YoshidaCBD, YoshidaSFF, delCampoGUESFF, Okuyama_2019}. For comparison, we also consider the unitary evolution of a single qubit in Sec.~\ref{sec:qubits}, finding exact results \emph{without} the need to average over an ensemble. Although we primarily use this result to benchmark calculations for the more general $N\!\times\!N$ case, it may also be relevant to developing strategies for error correction on NISQ devices.

Our results also reveal unexpected phenomenology. First, the fidelity of the depolarizing channel is \emph{nonmonotonic} in the noise strength. Consequently the ``typicality'' of the average channel (compared to typical instances of a noisy unitary) can be improved to some extent by \emph{increasing} the strength of the noise. Second, we observe that the ``strong-noise'' limit of the GUE channel does \emph{not} saturate the Haar-random limit, which leads to the maximally mixed state. We explain these counterintuitive results below, using the machinery of spectral form factors. We also comment that the existence of such unexpected phenomena even in the simple case of the GUE suggests that even more interesting behaviors may emerge in the context of noise drawn from more structured ensembles.

\section{Noisy single-qubit channels}
\label{sec:qubits}

We begin by considering arbitrary unitary operations on single qubits. In addition to providing a useful warm up to the more general noisy unitaries considered in Sec.~\ref{sec:channel def}, the single-qubit example admits exact solutions even without averaging over a particular ensemble of unitary operators, providing a useful benchmark for the more general case $N \geq 2$.

We first note that a single-qubit observable (or density matrix) $A$ can be decomposed onto the Pauli operators according to 
\begin{equation}
\label{eq:Pauli decomp}
    A = \frac{1}{2} \tr (A) \hspace{0.4mm} \ident + \bvec{a} \cdot \bvec{\sigma} \, , ~
\end{equation}
where the three-vector $\bvec{a}$ has coefficients $a_\nu  = \tr ( A \hspace{0.4mm} \sigma^\nu )/2$, and $\bvec{\sigma} = (\sigma^x,\sigma^y,\sigma^z)$ is a three-vector of Pauli matrices. 

Suppose that $A$ \eqref{eq:Pauli decomp} evolves in the Heisenberg picture under some single-qubit Hamiltonian $G$ of the form \eqref{eq:Pauli decomp}, i.e.,
\begin{equation}
\label{eq:qubit channel}
     A_{\bvec{g}} (t) = \ee^{\ii t \bvec{g} \cdot \bvec{\sigma}} A \hspace{0.4mm} \ee^{-\ii t \bvec{g} \cdot \bvec{\sigma}}  , ~~
\end{equation}
where the part of $G$ proportional to the identity acts trivially, and to capture the Schr\"odinger evolution of density matrices, one need only replace $t \to -t$. Using Euler's formula for quaternions, the unitary evolution operator can be written
\begin{equation}
\label{eq:Euler}
    U_G (t) = \ee^{-\ii t \bvec{g} \cdot \bvec{\sigma}} = \cos ( g t ) \ident - \ii \sin ( gt ) \hspace{0.4mm} \hat{\bvec{g}}\cdot \bvec{\sigma} \, ,~~ 
\end{equation}
where $g \coloneq \abs{\bvec{g}}$ is the magnitude of $\bvec{g}$ and $\hat{\bvec{g}} \equiv \bvec{g}/g$ is the associated \emph{unit} vector. Using \eqref{eq:Euler} and various properties of the Pauli matrices, we can rewrite $A_{\bvec{g}} (t)$ \eqref{eq:qubit channel} as
\begin{align}
    A_{\bvec{g}} (t) &= \frac{\tr(A)}{2} \ident + \cos (2gt) \hspace{0.4mm} \bvec{a} \cdot \bvec{\sigma} -\frac{\sin (2gt)}{g} ( \bvec{g} \times \bvec{a} ) \cdot \bvec{\sigma} \notag \\
    &\quad + \frac{2 \sin^2 (gt)}{g^2} ( \bvec{g} \cdot \bvec{a} ) ( \bvec{g} \cdot \bvec{\sigma} ) \, ,~\label{eq:A(t) Pauli}
\end{align}
which is exact for any Hermitian $A$ and $G$ and any time $t$.

We are primarily interested in the \emph{average} of the unitary channel \eqref{eq:qubit channel} over an RMT ensemble $\mathcal{E}$, captured by
\begin{equation}
    \label{eq:qubit avg channel}
    \Avg( A, t ) = \expec_{\mathcal{E}} [ A(t) ] \coloneq \int \thed^3 \bvec{g} \,  A_{\bvec{g}} (t) \,  P_{\mathcal{E}} ( \bvec{g} )\, ,~~
\end{equation}
where $P_{\mathcal{E}} ( \bvec{g} )$ is the probability density function (PDF) for $G = \bvec{g} \cdot \bvec{\sigma} \in \mathcal{E}$. Note that the trace part of $G$ is irrelevant to the channel \eqref{eq:qubit channel}. It is also straightforward to show that the map $A \mapsto \expval{A(t)}$ \eqref{eq:qubit avg channel} is completely positive and trace preserving (CPTP), and thus a valid quantum channel \cite{QC_book}.

In particular, we restrict to the GUE~\cite{mehta2004random, Livan_2018, eynard2018random, YoshidaCBD, TaoRMT, Brezin:2016eax}, which we review for completeness in App.~\ref{app:GUE}. In the case of a single-qubit system, the components $g_0 = \tr (G)/2$ and the three Pauli components $\bvec{g}$ \eqref{eq:Pauli decomp} are real valued and drawn from the normal distribution $\mathcal{N}_{\Reals} (0,\sigma^2/2)$. Ignoring the trace component $g_0$, as it does not contribute to the unitary channel \eqref{eq:qubit channel}, we find that
\begin{equation}
    \label{eq:P(G) qubit GUE}
    P_{\text{GUE}} (\bvec{g},\sigma) = \frac{1}{\pi^{3/2}\, \sigma^3 } \ee^{-\bvec{g}^2 / \sigma^2} \, , ~~ 
\end{equation}
which follows from the general relation for $N\! \times \! N$ matrices \eqref{eq:P of G GUE} in App.~\ref{app:GUE}. Using the PDF \eqref{eq:P(G) qubit GUE}, the average of the unitary channel $A_{\bvec{g}}(t)$ \eqref{eq:qubit channel} over the GUE leads to
\begin{equation}
\label{eq:qubit onefold channel 1}
    \Avg( A,t)   = \frac{1}{2} \tr (A) \ident + h (\sigma t) \, \bvec{a} \cdot \bvec{\sigma} \, ,~~ 
\end{equation}
where we have implicitly defined the function $h(z)$. Importantly, we note that the variance $\sigma$ only appears in the product $\sigma t$;  without loss of generality, we fix $\sigma = 1/\sqrt{2}$ in the remainder (see App.~\ref{app:GUE} for details). To facilitate comparison to results for $N>2$,  we rewrite the GUE-averaged channel $\expval{A_{\bvec{g}}(t)}$ \eqref{eq:qubit onefold channel 1} as 
\begin{figure}[t!]
    \centering
    \includegraphics[width=\linewidth]{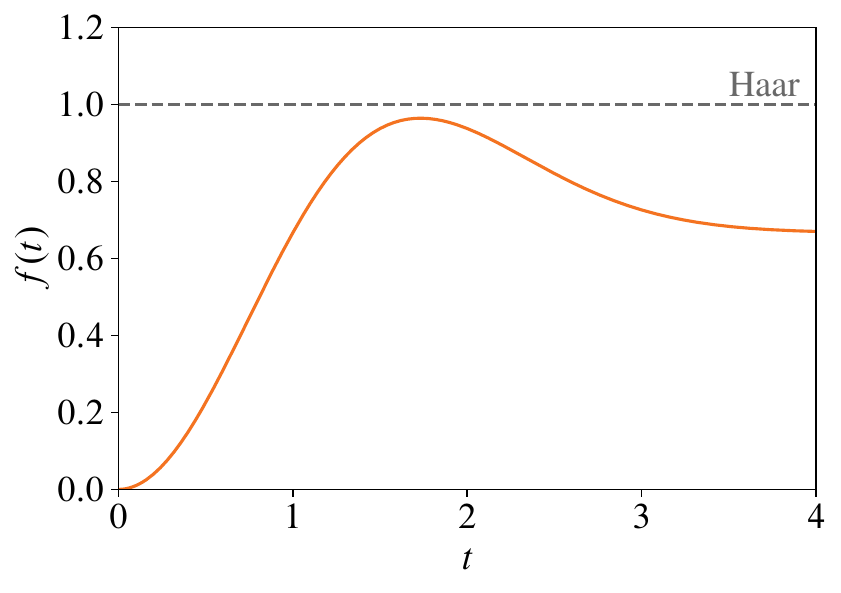}
    \caption{The amplitude $f(t)$ of the term $\tr(A) \ident /2$ in the GUE-averaged single-qubit channel $\expval{A_{\bvec{g}}(t)}$~\eqref{eq:qubit onefold channel}. For comparison, the dashed line indicates the Haar-random value $f(t)=1$, which is never realized by $\expval{A_{\bvec{g}}(t)}$. For $t \gtrsim 4$, $f(t)$ saturates to $2/3$.{\color{white} + AI. }}
    \label{fig:qubit-avg}
\end{figure}
\begin{align}
\label{eq:qubit onefold channel}
  \Avg( A,t )  &= \left[ 1 - f (t) \right] A + f( t) \frac{\tr (A) }{2} \hspace{0.4mm} \ident \, .~
\end{align}
which is parameterized by the single function $f(t)$, defined by
\begin{equation}
    f(t) \coloneq \frac{2}{3} \left[ 1- \ee^{-t^2/2}  \left(1-t^2 \right) \right] \, ,~~
    \label{eq:f2 of t}
\end{equation}
for qubits and for the choice $\sigma = 1/\sqrt{2}$ that we consider herein. 
In Fig.~\ref{fig:qubit-avg}, we plot the amplitude $f(t)$ that appears in the average channel \eqref{eq:qubit onefold channel}. We note that $f(t)$ is nonmonotonic in $t$ and never saturates the Haar-random limit $f(t) = 1$. In Sec.~\ref{subsec:average chan form}, we generalize $f(t)$~\eqref{eq:f2 of t} to arbitrary $N \geq 2$.

When applied to density matrices $\rho$, the GUE-averaged channel $\rho \mapsto \expval{\rho_{\bvec{g}}(t)}$ \eqref{eq:qubit onefold channel} realizes a \emph{depolarizing channel}~\cite{QC_book} with a $t$-dependent polarization parameter $p$, i.e.,
\begin{equation}
    \label{eq:depolarizing}
    \Phi_{\text{dp}} (\rho) = (1-p) \rho + \frac{p}{N} \ident \, , ~~
\end{equation}
where $p = f(t)$ \eqref{eq:f2 of t}, and we used the fact that $\tr (\rho)=1 $. At $t=0$, the $\rho$ term has coefficient one, while the $\ident$ term---corresponding to the maximally mixed state $\rho_\infty = \ident / N$---has coefficient zero. At early times, the coefficient of the maximally mixed state grows like $t^2 + \Order{t^4}$, and is maximized when $t = \sqrt{3}$, where $p_{\text{max}} \approx 0.96 $. Importantly, this coefficient is always less than one, meaning that the GUE-averaged channel $\expval{\rho_{\bvec{g}}(t)}_{\text{GUE}}$ \eqref{eq:qubit onefold channel} never realizes the Haar-random limit $\expval{\rho_{\bvec{g}}(t)}_{\text{Haar}} = \ident/2$, due to the additional structure of the GUE compared to the uniform Haar measure \cite{BnB, YoshidaSFF}. Moreover, $p = f (t)$ is a \emph{nonmonotonic} function of $t$ (see Fig.~\ref{fig:qubit-avg}), which decreases for $t > \sqrt{3}$, saturating to $f (t) = 2/3$ for $t \gtrsim 4$. However, we are primarily interested in the limit $t \ll 1$, corresponding to \emph{weak} noise.

Finally, to probe the extent to which the average GUE channel $\expval{ A_{\bvec{g}}(t) }$ \eqref{eq:qubit onefold channel} captures \emph{typical} values of the time-evolved operator $A_{\bvec{g}}(t)$ \eqref{eq:A(t) Pauli}, we evaluate the variance of $A_{\bvec{g}}(t)$ for $G$ drawn from the GUE~\cite{mehta2004random, Livan_2018, eynard2018random, YoshidaCBD, TaoRMT, Brezin:2016eax}. In particular, we define the quantity
\begin{align}
    \var(A,t) &\coloneq \expec_{\mathcal{E}} \left[ A^{*}_{\bvec{g}}(t) \otimes A^{\vps}_{\bvec{g}} (t)\right] -  \expec_{\mathcal{E}} \left[ A_{\bvec{g}}(t) \right]^{\otimes 2} \label{eq:VarA def qubit} \\
    &\; = \frac{7 f (t) - 3 f (2 t) - 5 f (t)^2}{5} \left( \bvec{a} \cdot \bvec{\sigma} \right)^* \otimes \left( \bvec{a} \cdot \bvec{\sigma} \right) \notag \\
    &~~+ \frac{ f ( t) + f (2 t)}{5} \abs{a}^2 \sum\limits_{n=1}^3 \left( \sigma^n \right)^* \otimes \sigma^n \, ,~~\label{eq:VarA result qubit}
\end{align}
which we use to extract the variance of the individual matrix elements of $A_{\bvec{g}}(t)$ \eqref{eq:A(t) Pauli}. We note that the latter operator can be written in terms of the SWAP operator $\mathcal{S} = \sum \hspace{0mm}_{m,n} \ket{nm}\bra{mn}$, which is relevant to the general $N$ case in Sec.~\ref{subsec:variance chan form}.

We use $\var (A,t)$~\eqref{eq:VarA result qubit} to compute the variance of individual matrix elements of $A_{\bvec{g}}(t)$~\eqref{eq:qubit channel} by evaluating
\begin{align}
    \var(A_{mn},t) &\coloneq \expval{ \abs{A_{mn} (t)}^2} - \abs{\expval{A_{mn}(t)}}^2 \notag \\
    &~= \tr \Big( \var(A,t) \, \BKop{n}{m}^{\otimes 2} \Big)  \label{eq:VarAmn def qubit}  \, ,~~
\end{align}
so that the two diagonal elements $A_{nn}$ have variance
\begin{align}
    \var(A_{nn},t) &= \frac{7 f (t) - 3 f (2t) - 5 f (t)^2}{5} a_z^2  \notag \\
    &~+ \frac{f (t) + f (2t) }{5} \abs{a}^2 \label{eq:VarAdiag def qubit} \, , ~~
\end{align}
while the two off-diagonal elements have variance
\begin{align}
    \var(A_{mn},t) &= -\frac{7 f (t) - 3 f (2t) - 5 f (t)^2}{5} a_z^2  \notag \\
    &~+ \frac{9 f (t) - f (2t) -5 f (t)^2}{5} \abs{a}^2 \label{eq:qubit offdiag matel var} \, , ~~
\end{align}
for $m\neq n$. Note that $a_z$ and $a = \abs{\bvec{a}}$ refer to the Pauli components of $A$ \eqref{eq:Pauli decomp} \emph{prior} to evolution under the Gaussian Hamiltonian $A \mapsto A_{\bvec{g}}(t)$~\eqref{eq:qubit channel} and the trace part $a_0 = \tr (A)/2$ does not contribute to the variances---intuitively, this is because the identity piece is preserved by the channel~\eqref{eq:qubit avg channel}.

\begin{figure}[t!]
    \centering
    \includegraphics[width=\linewidth]{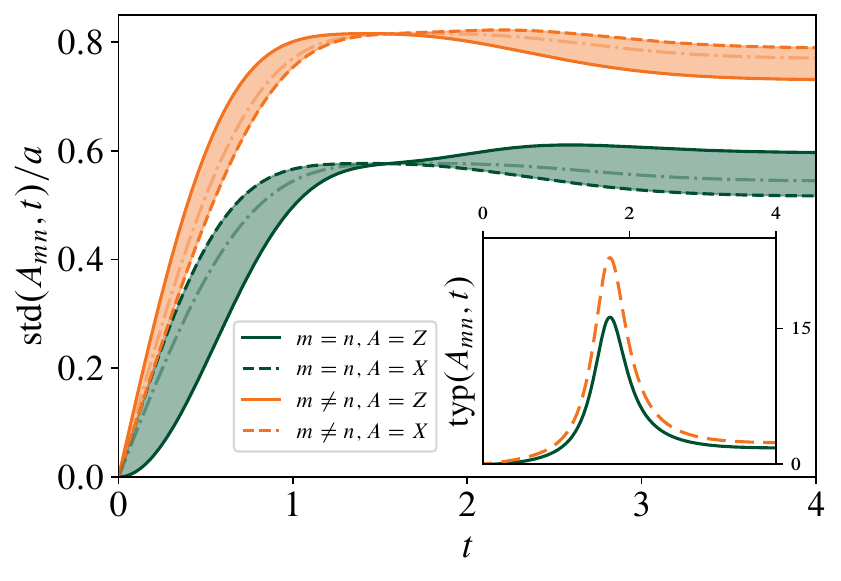}
    \caption{Standard deviation of matrix elements $A_{mn}(t)$ of $A_G(t)$~\eqref{eq:qubit channel} for an initial operator $A= \bvec{a} \cdot \bvec{\sigma}$, in units of $a = \abs{\bvec{a}}$. Diagonal elements ($m=n$) appear in green and off-diagonal elements ($m \neq n$) appear in orange. The choices $A=Z$ (solid) and $A=X$ (dashed) bound the standard deviation for a general operator, corresponding to the shaded region; the dash-dotted curves represent the same standard deviations with $A$ averaged over the GUE, appropriately normalized (see also Sec.~\ref{subsec:variance chan form}). \textbf{Inset}: Plot of a ``typicality'' diagnostic~\eqref{eq:qubit typ def}, which is the ratio of the standard deviation to the mean. We find that $\operatorname{typ}(A_{mn},t) \ll 1$ is small when $t \lesssim 1$, meaning that the average~\eqref{eq:qubit onefold channel} well approximates a typical instance of the noisy unitary~\eqref{eq:qubit channel}. For $t \gtrsim 3$, we find a somewhat typical regime with and $\operatorname{typ}(A_{mn},t) \sim 1$.}
    \label{fig:qubit-var}
\end{figure}

We plot the standard deviations of the matrix elements $A_{nn}$~\eqref{eq:VarAdiag def qubit} and $A_{mn}$~\eqref{eq:qubit offdiag matel var} in Fig.~\ref{fig:qubit-var} in units of $a = \abs{\bvec{a}}$. The standard deviation of matrix elements for generic single-qubit operators lie in the shaded regions in Fig.~\ref{fig:qubit-var}; the upper and lower bounds correspond to: (\emph{i}) $a_{\smash z}^{\smash 2}=1$ with $a^{\smash \vps}_{\smash x}=a_{\smash y}^{\smash \vps}=0$ and (\emph{ii}) $a_{\smash z}^{\smash \vps}=0$ with $a_{\smash x}^{\smash 2}+a_{\smash y}^{\smash 2} = 1$. These limiting cases are captured by the choices $A=Z$ (solid) and $A=X$ (dashed), respectively. We also plot the standard deviation with $A$ averaged over the $2 \! \times \! 2$ GUE with variance $\sigma_A^2=2/3$, which ensures that $\smash{\expval{\abs{a}^2}_{A \in \text{GUE}} = 1}$. Using these quantities, we also define the following measure of ``typicality'' of the noisy channel~\eqref{eq:qubit onefold channel}, 
\begin{align}
    \operatorname{typ} \left( A_{mn}, t \right) &\coloneq \left[ \frac{\tr \left[ \operatorname{var} (A,t) \BKop{n}{m}^{\otimes 2} \right]}{\tr \left[ \operatorname{avg} (A,t)^{\otimes 2} \BKop{n}{m}^{\otimes 2} \right]} \right]^{1/2} \notag \\
    &\eqcolon \frac{\operatorname{std} \left( A_{mn}, t \right)}{\operatorname{avg} \left( A_{mn}, t \right)} \, ,~~\label{eq:qubit typ def}
\end{align}
which is small when the average channel~\eqref{eq:qubit onefold channel} is close to a ``typical'' instance of a noisy unitary~\eqref{eq:qubit channel}.
Indeed, if~\eqref{eq:qubit typ def} is much larger than one, the number of samples required to produce fractional error $\epsilon$ scales as $\sim (\operatorname{typ} / \epsilon)^2$ by the central limit theorem.
This measure of typicality~\eqref{eq:qubit typ def} is plotted in the inset of Fig.~\ref{fig:qubit-var}, where we assume $A$ to be traceless for convenience; to ensure that the mean is nonzero, we further take $A=Z$ for the typicality of diagonal elements, and take $A=X$ for the typicality off-diagonal elements, since we have that
\begin{subequations}
\label{eq:qubit matel means}
\begin{align}
    \operatorname{avg}(A_{nn},t) &= \abs{ a^{\vps}_0 + (-1)^n \left[ 1 - f(t) \right]  a^{\vps}_z}  \label{eq:qubit matel diag mean} \\
    \operatorname{avg}(A_{mn},t) &= \left[1 - f(t) \right] \sqrt{a_x^2 + a_y^2} \label{eq:qubit matel offdiag mean} \, ,~~
\end{align}
\end{subequations}
for $i \neq j$ and a generic initial operator $A = \bvec{a} \cdot \bvec{\sigma}$~\eqref{eq:Pauli decomp}, where we use the computational ($Z$) basis such that $Z \ket{j}=(-1)^j \ket{j}$.

We conclude with several remarks on Fig.~\ref{fig:qubit-var}. First, we note that the average~\eqref{eq:qubit onefold channel}, variance~\eqref{eq:VarA result qubit}, and  typicality~\eqref{eq:qubit typ def} are all \emph{nonmonotonic} functions of $t$. Second, we note that all three quantities saturate to late-time values, which do not saturate the Haar-random limit (i.e., $f(t) < 1$ for all $t$). We also note that the diagonal elements have lower standard deviation because the identity term $a_0 \ident$~\eqref{eq:Pauli decomp} does not vary over the ensemble. Finally, we observe that the typicality is small---meaning that the average channel~\eqref{eq:qubit onefold channel} reflects a typical instance of noisy unitary evolution~\eqref{eq:qubit channel}---for $t \lesssim 1$. Somewhat surprisingly, when $t \gtrsim 3$, we find a regime of \emph{somewhat} typical behavior in which the typicality of matrix elements~\eqref{eq:qubit typ def} is exactly $1$.

\section{Generic noisy channels}
\label{sec:channel def}

We now consider noisy evolution on an $N$-dimensional Hilbert space $\Hilbert = \Comps^N$.  As in the qubit case, we evaluate the unitary evolution of operators $A$ in the Heisenberg picture,
\begin{equation}
\label{eq:noisyOp}
    A_G (t) = U^{\dagger}_G (t) \hspace{0.2mm} A \hspace{0.2mm}  U^{\vpd}_G (t) \equiv \ee^{\ii G t} A  \hspace{0.2mm} \ee^{-\ii G t} \hspace{0.2mm} ,~~
\end{equation}
where noise is introduced by drawing the $N \! \times \! N$ Hermitian operator $G$ from some ensemble $\mathcal{E}$, and the analogous evolution of density matrices in the Schr\"odinger picture recovers by replacing the operator $A$ with a density matrix $\rho$ and taking $t \to -t$ (this may act trivially, as is the case for the GUE).

Of particular interest is the nonunitary quantum \emph{channel} corresponding to the \emph{average} of the $A_G (t)$ \eqref{eq:noisyOp} over statistically similar ``Hamiltonians'' $G \in \mathcal{E}$, sampled with probability density $P_\mathcal{E}(G)$. Using standard RMT ensembles allows for the analytical evaluation of the average of $A_G(t)$ \eqref{eq:noisyOp}, as well as higher moments like the variance of matrix elements \eqref{eq:VarAmn def qubit}.

Below, we take $\mathcal{E}$ to be the GUE \cite{mehta2004random, Livan_2018, eynard2018random, YoshidaCBD, TaoRMT, Brezin:2016eax} with variance $\sigma^2 = 1/2$, as discussed in App.~\ref{app:GUE}. The components $G_{ij}$ of an element $G$ of the GUE are then sampled from the PDF
\begin{equation}
\label{eq:Gij prob}
    P_{\text{GUE}}(G_{ij}) = \mathcal{N}_{ij} (0,1/2) \, ,~~
\end{equation}
for $i \leq j$, where $\mathcal{N}_{ij}(\mu,\sigma^2)$ is the \emph{real} normal distribution for $i=j$, and the \emph{complex} normal distribution when $i \neq j$ \cite{mehta2004random, Livan_2018, eynard2018random, YoshidaCBD, TaoRMT, Brezin:2016eax}. 
The corresponding PDF for the matrix $G$ is
\begin{equation}
    \label{eq:GUE P(G)}
    P_{\text{GUE}}(G) = \frac{1}{\Omega} \hspace{0.2mm} \ee^{- \tr ( G^2 )} \, ,~~
\end{equation}
where $G$ is Hermitian and $\Omega$ is chosen to ensure normalization.

The Hilbert space $\Hilbert = \Comps^N$ may describe a many-body system (e.g., $n$ qubits, so that $N = 2^n$). The Hilbert space $\Hilbert$ may also include ancillary (or ``environmental'') degrees of freedom, which may be traced out or projected onto a particular state, where one expects to evolve operators $A$ that act trivially on the environment. Alternatively, $\Hilbert$ may be a \emph{subset} of a larger Hilbert space, in which case one should decompose the full operator onto a basis of Kronecker products of local operators (e.g., the Pauli strings), and apply the evolution \eqref{eq:noisyOp} to the part of each basis string that acts on $\Hilbert$. For example, one can apply a single-qubit channel to the operator $X \otimes X + Y \otimes Y$ by evaluating $X_G (t) \otimes X + Y_G (t) \otimes Y$. Additionally, one can introduce block structure in $G$---corresponding to symmetries and/or kinetic constraints---by including projectors, as is common in the literature on quantum chaos~\cite{U1FRUC, ConstrainedRUC, MIPT}.

Of particular interest is the ability to tune the strength of the noise continuously to zero, so that $U_G = \ident$ acts trivially for vanishing noise strength. In general, one expects the noise ``strength'' to be captured by the \emph{variance} $\sigma$ of the PDF $P_\mathcal{E}(G)$. However, as we saw in the case of qubits, the average $\expval{A_G(t,\sigma)}$ depends on $\sigma$ and $t$ only through the product $\sigma t$. As discussed in Sec.~\ref{sec:qubits} and App.~\ref{app:GUE}, we take $\sigma = 1/\sqrt{2}$ without loss of generality in the remainder, and parameterize the noise strength via $t$ alone. Lastly, we comment that the methods below extend to \emph{any} rotation-invariant ensemble $\mathcal{E}$~\cite{Vinayak_2012, _nidari__2011}, including the three Gaussian~\cite{mehta2004random, Livan_2018, eynard2018random, YoshidaCBD, TaoRMT, Brezin:2016eax, GOEfuture} and Wishart-Laguerre~\cite{Wishart, Livan_2018} ensembles, and the Altland-Zirnbauer~\cite{AZensembles} ensembles.

\subsection{Ensemble-averaging procedure}
\label{subsec:eigen pdf}

We now explain the procedure for ensemble averaging, which holds for generic random-matrix ensembles $\mathcal{E}$ that are invariant under some ``rotation'' group $\mathsf{R}$ in the sense that $P_\mathcal{E} (G) = P_\mathcal{E}( V G V^\dagger )$ for all $V \in \mathsf{R}$. We elucidate the notion of rotation invariance in App.~\ref{app:rotInv}, and further require that the matrix $V_G$ that diagonalizes $G$ be an element of $\mathsf{R}$. For example, for the Gaussian orthogonal ensemble (GOE), $\mathsf{R}=\Orth{N}$; for the Gaussian unitary ensemble (GUE), $\mathsf{R}=\Unitary{N}$; and for the Gaussian symplectic ensemble (GSE), $\mathsf{R}=\Symp{N}$~\cite{mehta2004random}. Other RMT ensembles---including the Wishart-Laguerre~\cite{Wishart, Livan_2018} and Altland-Zirnbauer~\cite{AZensembles} ensembles---also satisfy rotation invariance, and are amenable to the analytic procedure below.

The noisy evolution of an observable $A \mapsto A_G(t)$ \eqref{eq:noisyOp} under a Hamiltonian $G$ drawn from a rotation-invariant ensemble $\mathcal{E}$ is fully characterized by the various \emph{moments} of $A_G(t)$ \eqref{eq:noisyOp}. The $k$th moment of $A_G(t)$ \eqref{eq:noisyOp} is given by the $k$-fold channel
\begin{align}
\label{eq:nfold channel}
    \varphi^{(k)}_{\mathcal{E}} (A,t) &\coloneq \expec_{\mathcal{E}} \left[ A_G (t)^{\otimes k} \right] \notag \\
    &\hphantom{:}= \int_{\mathcal{E}}  \thed G \, P_{\mathcal{E}} (G) \, \left( \ee^{\ii G t} A \, \ee^{-\ii G t} \right)^{\otimes k}\, ,
\end{align}
where this channel can also be applied to the Kronecker product of $k$ distinct operators $A_i$, as in the case of $\varphi^{(2)}(A,B;t)$~\eqref{eq:Var Chan form}.

We are particularly interested in the GUE, in which case $\thed G \equiv \prod_{i\leq j} \thed \Re{G_{ij}}\prod_{i <j} \thed \Im{G_{ij}}$ is the measure \eqref{eq:Gij prob} for the $N^2$ independent, real parameters that uniquely specify $G$. These correspond to $N$ diagonal elements $G_{ii} \in \Reals$ and $N(N-1)/2$ upper-triangular elements $G_{ij} \in \Comps$ for $i<j$, which have independent real and imaginary parts (the lower-triangular elements of $G$ are then fixed by the fact that $G$ is Hermitian)~\cite{mehta2004random, Livan_2018, eynard2018random, YoshidaCBD, TaoRMT, Brezin:2016eax}. Further details appear in App.~\ref{app:GUE}, and we comment that RMT ensembles other than the GUE generally involve further restrictions on the matrix elements of $G$.

We now evaluate the moments of $A_G(t)$ \eqref{eq:noisyOp}. Of particular interest are the \emph{average} of $A_G(t)$, given by
\begin{equation}
\label{eq:Avg Chan Def}
    \Avg ( A,t ) = \varphi^{(1)}_{\mathcal{E}} (A,t) = \int_{\mathcal{E}}  \thed G \, P_{\mathcal{E}} (G) \, \ee^{\ii G t} A \, \ee^{-\ii G t} \, ,~~
\end{equation}
along with the \emph{variance} of $A_G(t)$, given by
\begin{equation}
\label{eq:Var Chan Def}
    \var ( A,t ) = \varphi^{(2)}_{\mathcal{E}} (A^*,A;t) - \varphi^{(1)}_{\mathcal{E}} (A,t)^2 ,~~
\end{equation}
which can be used to diagnose, e.g., typicality of the mean~\eqref{eq:Avg Chan Def}.

We now explain how to use \emph{rotation invariance} of the PDF $P_{\text{GUE}}(G)$ \eqref{eq:GUE P(G)} to simplify the evaluation of the $k$-fold channels \eqref{eq:nfold channel}. Further details of this procedure appear in App.~\ref{app:rotInv}, and though we restrict to the GUE below, these results hold \emph{mutatis mutandis} for \emph{generic} rotation-invariant matrix ensembles $\mathcal{E}$. In particular, it is easy to check that 
\begin{equation}
    \label{eq:GUE rotation invariance}
    P_{\text{GUE}}( U^\dagger G \hspace{0.2mm} U) = P_{\text{GUE}} (G) ~,~~\forall \, U  \in \Unitary{N} \, .~
\end{equation}
Next, we use rotation invariance of the ensemble $\mathcal{E}$ to simplify the integrals over the independent elements of $G \in \mathcal{E}$ to integrals over the $N$ eigenvalues $\{ \lambda_n \}$ of $G$, and over the eigenvectors, which are given by the columns of a unitary matrix $V_{\smash G}$. This is accomplished using the factorization~\cite{mehta2004random}
\begin{equation}
    \label{eq:spectral pdf}
    P_{\mathcal{E}} (G) \, \thed G = \rho (\bvec{\lambda}) \, p (V^{\vpd}_{\smash G})  \hspace{0.2mm}  \thed \bvec{\lambda} \hspace{0.2mm} \thed V^{\vpd}_{\smash G} \, ,~~
\end{equation}
where $\bvec{\lambda}=(\lambda_1 , \dots , \lambda_n)$ is the $N$-component vector whose entries are the eigenvalues of $G$, and we use $\rho (\bvec{\lambda})$ to denote the joint eigenvalue PDF, also known as the $N$-point eigenvalue density function of $G$ (see App.~\ref{app:averaging} and Ref.~\citenum{mehta2004random} for additional details). The spectral decomposition \eqref{eq:spectral pdf} holds for \emph{any} rotation-invariant RMT ensemble, and simplifies calculations substantially \cite{Vinayak_2012}. With the GUE, both sides of \eqref{eq:spectral pdf} involve $N^2$ real parameters and the PDF $p (V^{\,}_{\smash G})$ for the diagonalizing unitary $V^{\,}_{\smash G}$ \eqref{eq:diagonalizing unitary VG} constitutes a measure over the unitary group $\Unitary{N}$. For other RMT ensembles, the number of unique parameters generally differs, as does the rotation group $\mathsf{R}$ from which $V^{\,}_{\smash G}$ is drawn  (e.g., $\mathsf{R} = \Orth{N}$ for the GOE).

Using rotation invariance, we now work out $p (V^{\,}_{\smash G})$~\eqref{eq:spectral pdf}. We first note that the map $G \mapsto U \hspace{0.2mm} G \hspace{0.2mm} U^\dagger$ that leaves $P_{\text{GUE}}(G)$ invariant is equivalent to sending $V^{\,}_{\smash G} \mapsto U \hspace{0.2mm} V^{\,}_{\smash G}$ \eqref{eq:G spec decomp}; hence, we have that $p (U \hspace{0.2mm} V^{\,}_{\smash G}) = p (V^{\,}_{\smash G})$ for all $U \in \Unitary{N}$. Moreover, because $V^{\,}_{\smash G}$ is an element of a \emph{compact} group, such ``left invariance'' immediately implies ``left-right invariance''---i.e., $p (U \hspace{0.2mm} V^{\,}_{\smash G} \hspace{0.2mm} W) = p (V^{\,}_{\smash G})$ for all $U,W \in \Unitary{N}$~\cite{Simon1995RepresentationsOF, Collins_2003, Collins_2006, Collins_2009, collins2021weingarten}. Such left-right invariance of $p (V^{\,}_{\smash G})$~\eqref{eq:spectral pdf} in turn implies that $p (V_G)$ is  the (uniform) \emph{Haar measure} over $\Unitary{N}$.
More generally, for ensembles $\mathcal{E}$ other than the GUE, we instead have that $p (V^{\,}_{\smash G})$~\eqref{eq:spectral pdf} is the uniform (Haar) measure over the corresponding rotation group $\mathsf{R}$ under which $\mathcal{E}$ is invariant.

Finally, the joint eigenvalue PDF $\rho (\bvec{\lambda})$ \eqref{eq:spectral pdf} is given by
\begin{equation}
    \label{eq:GaussianSpectralDensity}
    \rho (\bvec{\lambda}) = \frac{1}{\Omega} \, \ee^{- \sum_{n=1}^{N}\, \lambda^2_n} \, \prod\limits_{i<j} \, \abs{\lambda_i - \lambda_j}^2 \, ,~~
\end{equation}
where the ``partition function'' $\Omega$ ensures normalization, and the product over $\abs{\lambda_i - \lambda_j}^2$ results from the change of variables from $G$ to $\bvec{\lambda}$ and $V_G$; see App.~\ref{app:rotInv} for further detail.

In computing the moments $\varphi^{(k)}(A,t)$~\eqref{eq:nfold channel}, we first evaluate the Haar integrals over the eigenvectors $V_G$; the remaining integrals over the eigenvalues $\bvec{\lambda}$ results in various spectral form factors (SFFs)---i.e., Fourier transforms of the eigenvalue density function~\cite{CDLC1, U1FRUC, ConstrainedRUC, mehta2004random, guhr1998random, YoshidaSFF, ShenkerRMT, Vinayak_2012, MIPT, RMT_SFF, YoshidaCBD, delCampoGUESFF, Micklitz, Okuyama_2019, Brezin:2016eax, BohigasChaos, AmitQuantErgo}. Integrating $V_G$ \eqref{eq:diagonalizing unitary VG} over the Haar measure results in ``contractions'' between various indices of the $k$ copies of $A$, and between the $2k$ terms of the form $\ee^{\pm \ii t \lambda_n}$, where the latter take the form of Fourier transforms of $\rho (\bvec{\lambda})$. In Sec.~\ref{subsec:average chan form}, we write the average channel $\Avg(A,t)$ \eqref{eq:Avg Chan Def} in terms of the standard two-point SFF $\mathcal{R}_2(t)$; in Sec.~\ref{subsec:variance chan form}, we write the variance $\var(A,t)$ \eqref{eq:Var Chan form} in terms of four-, three-, and two-point SFFs; in Sec.~\ref{subsec:higher cum form}, we discuss the evaluation of higher moments (or cumulants) in terms of higher-point SFFs; and finally, the SFFs themselves are evaluated in Sec.~\ref{sec:SFF}. As a reminder, analogous derivations apply to other rotation-invariant ensembles, such as the GOE~\cite{GOEfuture, Vinayak_2012}.

\subsection{Average of the noisy channel}
\label{subsec:average chan form}

\begin{figure*}[t]
    \centering
    \includegraphics[width=\linewidth]{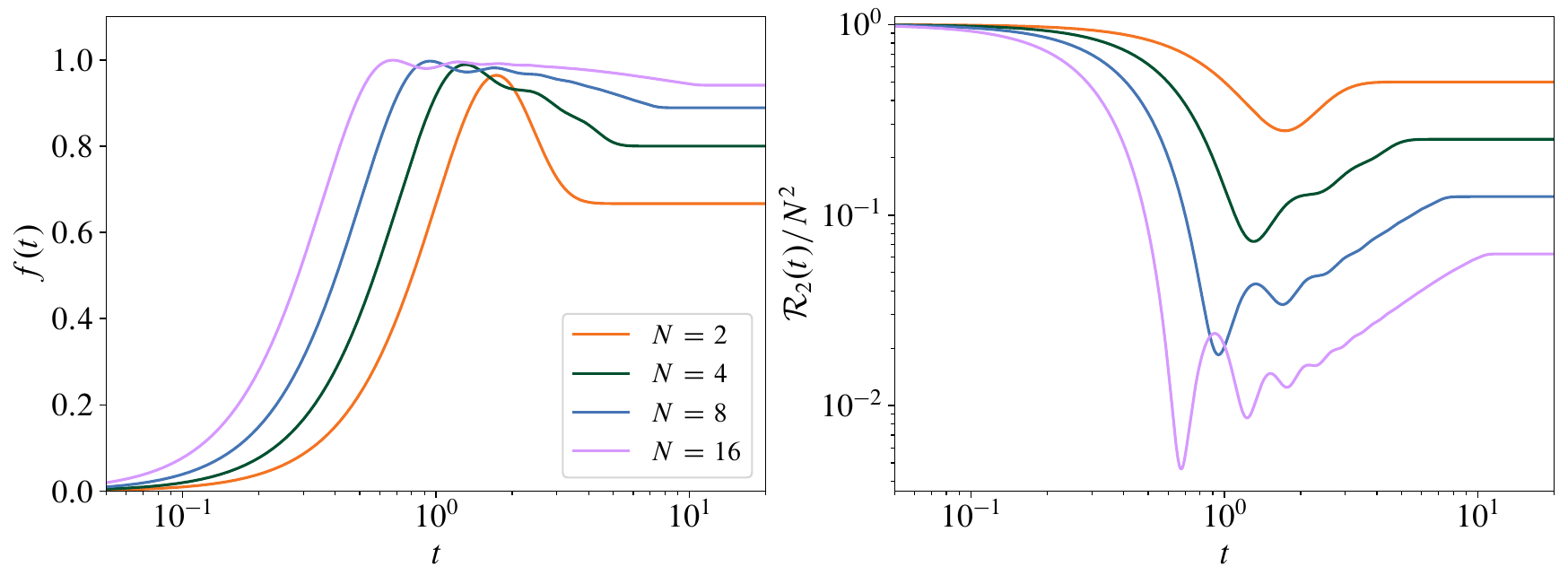}
    \caption{
    \textbf{Left}: Channel amplitude $f(t)$~\eqref{eq:f(t) def} for Hilbert-space dimensions $N=2^k$ for $k=1,2,3,4$, which captures the strength of the $\tr(A) \ident / N$ term in the average channel~\eqref{eq:Avg Noisy Chan two terms}. \textbf{Right}: The two-point SFF $\mathcal{R}_2(t)$~\eqref{eq:SFF main} normalized by its $t=0$ value of $N^2$, for the same values of $N=2^k$. 
    }
    \label{fig:SFFsFigure}
\end{figure*}

Here we present the average channel $\Avg ( A,t ) = \varphi^{(1)}_\mathcal{E}(A,t)$ \eqref{eq:Avg Chan Def} for a generic ensemble $\mathcal{E}$ that is invariant under rotations in $\mathsf{R}=\Unitary{N}$, such as the GUE~\cite{mehta2004random, Livan_2018, eynard2018random, YoshidaCBD, TaoRMT, Brezin:2016eax}. A more detailed derivation appears in App.~\ref{app:k=1 avg}. In particular, we have that
\begin{align}
\label{eq:Avg Chan form}
    \Avg ( A,t ) 
    &= \int \thed \bvec{\lambda} \, \rho (\bvec{\lambda}) \, \sum\limits_{m,n=1}^{N} \, \ee^{\ii t (\lambda_m - \lambda_n)} \; \times \notag \\
    &~~~ \int_{\mathsf{R}} \thed V \, p (V) \,  \, V \BKop{m}{m} V^\dagger \, A \, V \, \BKop{n}{n} V^\dagger \, ,~
\end{align}
where we omit the ``$G$'' subscript from $V_G$ \eqref{eq:diagonalizing unitary VG} for convenience of presentation. Integrating $V$ and $V^\dagger$ over $\mathsf{R}=\Unitary{N}$ with Haar measure leads to~\cite{BnB, Collins_2003},
\begin{align}
\label{eq:Avg Chan intermediate}
    \Avg ( A,t )  &= \int \thed \bvec{\lambda} \, \rho (\bvec{\lambda}) \sum_{m,n=1}^{N} \, \ee^{\ii t (\lambda_m - \lambda_n)} \, \Phi_{mn} (A) \, ,~
\end{align}
where the operator content is captured by
\begin{align}
\label{eq:Onefold Haar outcome}
    \Phi_{mn} (A) &= \frac{N - \kron{mn}}{N(N^2-1)} A + \frac{(N \, \kron{mn} - 1) \,\tr(A)}{N(N^2-1)} \ident \, ,~~
\end{align}
as we derive explicitly in App.~\ref{app:k=1 avg}. Integrating over the eigenvalues $\bvec{\lambda}$, the average channel \eqref{eq:Avg Chan intermediate} becomes
\begin{align}
    \label{eq:Avg Noisy Chan two terms}
    \Avg ( A,t )  = [1-f (t)] \, A + \frac{1}{N} \, f (t) \, \tr(A) \, \ident \, ,~~
\end{align}
which takes the form of a \emph{depolarizing channel}~\cite{QC_book}. The time-dependent depolarizing parameter is given by
\begin{equation}
\label{eq:f(t) def}
    f (t) \coloneq \frac{N^2-\mathcal{R}_2 (t)}{N^2-1} \, ,~~
\end{equation}
which agrees with the results for qubits~\eqref{eq:f2 of t} when $N=2$, and where $\mathcal{R}_2(t)$ is the two-point SFF~\cite{RMT_SFF, YoshidaSFF, CDLC1,  U1FRUC, ConstrainedRUC, MIPT, ShenkerRMT} for $\mathcal{E}$, i.e.,
\begin{align}
\label{eq:SFF main}
    \mathcal{R}_2(t) \coloneq \mathbb{E}_\mathcal{E} \, \sum\limits_{m,n=1}^{N} \, \ee^{\ii (\lambda_m-\lambda_n) t}
    \, ,
\end{align}
where $\lambda_{m,n}$ are the eigenvalues of the Hamiltonian $G$~\eqref{eq:G spec decomp}, so that $\mathcal{R}_2(t)$ \eqref{eq:SFF main} is the Fourier transform of the pair density function of energy eigenvalues of $G$. We calculate this quantity explicitly for the GUE in Sec.~\ref{subsec:SFF 2} for any finite $N$. We also note that, because $f(t)$~\eqref{eq:f(t) def} is an even function, both the Heisenberg evolution of operators $A$ and the Schr\"odinger evolution of density matrices $\rho$ are described by $\Avg ( A,t )$~\eqref{eq:Avg Noisy Chan two terms}, and one need only replace $A \to \rho$ in the latter case.

In Fig.~\ref{fig:SFFsFigure}, we plot both $f(t)$~\eqref{eq:f(t) def} and the SFF $\mathcal{R}_2(t)$~\eqref{eq:SFF main} for the GUE for several values of $N=2^k$ (i.e., corresponding to a system of $k$ qubits). Of particular interest is the fact that both $f(t)$~\eqref{eq:f(t) def} and $\mathcal{R}_2(t)$~\eqref{eq:SFF main} are nonmonotonic and even functions of $t$. We observe that $f(t)$ vanishes only at $t=0$, never saturates the Haar-random limit $f(t)=1$ (see also Fig.~\ref{fig:qubit-avg}), and asymptotes to $f(t)=N/(N+1)$ for $t \gtrsim N$. This can be attributed to properties of the SFF~\eqref{eq:SFF main}, which has value $N^2$ at $t=0$, plummets exponentially during an initial ``dip'' regime, followed by oscillations until the start of a hallmark linear ``ramp'' regime with $\mathcal{R}_2(t) \sim \abs{t}$, before reaching the ``plateau'' regime with $\mathcal{R}_2(t) = N$ for all times $t \geq N$~\cite{RMT_SFF, YoshidaSFF, CDLC1,  U1FRUC, ConstrainedRUC, MIPT, ShenkerRMT}. We note that the ramp is a signature of chaotic level repulsion, the start of the ramp is known as the ``Thouless time,'' and the start of the plateau as the ``Heisenberg time''~\cite{RMT_SFF, YoshidaSFF, CDLC1,  U1FRUC, ConstrainedRUC, MIPT, ShenkerRMT}.

\subsection{Variance of the noisy channel}
\label{subsec:variance chan form}

The average channel~\eqref{eq:Avg Noisy Chan two terms} is the quantity most relevant to experiments on noisy quantum devices. However, it is valuable to diagnose the extent to which that average $\operatorname{avg}(A,t)$~\eqref{eq:Avg Noisy Chan two terms} reflects a \emph{typical} instance $A_G (t)$~\eqref{eq:noisyOp} of noisy unitary evolution. To do so, it is valuable to compare the variance of the channel~\eqref{eq:Var Chan Def} with the mean~\eqref{eq:Avg Chan Def}. As in Sec.~\ref{sec:qubits}, we make this comparison at the level of matrix elements $A_{mn} (t)$, and, for convenience, we draw the observable $A$ from the GUE with variance $\sigma_A^2$. 

\begin{figure*}[t]
    \centering
    \includegraphics[width=\linewidth]{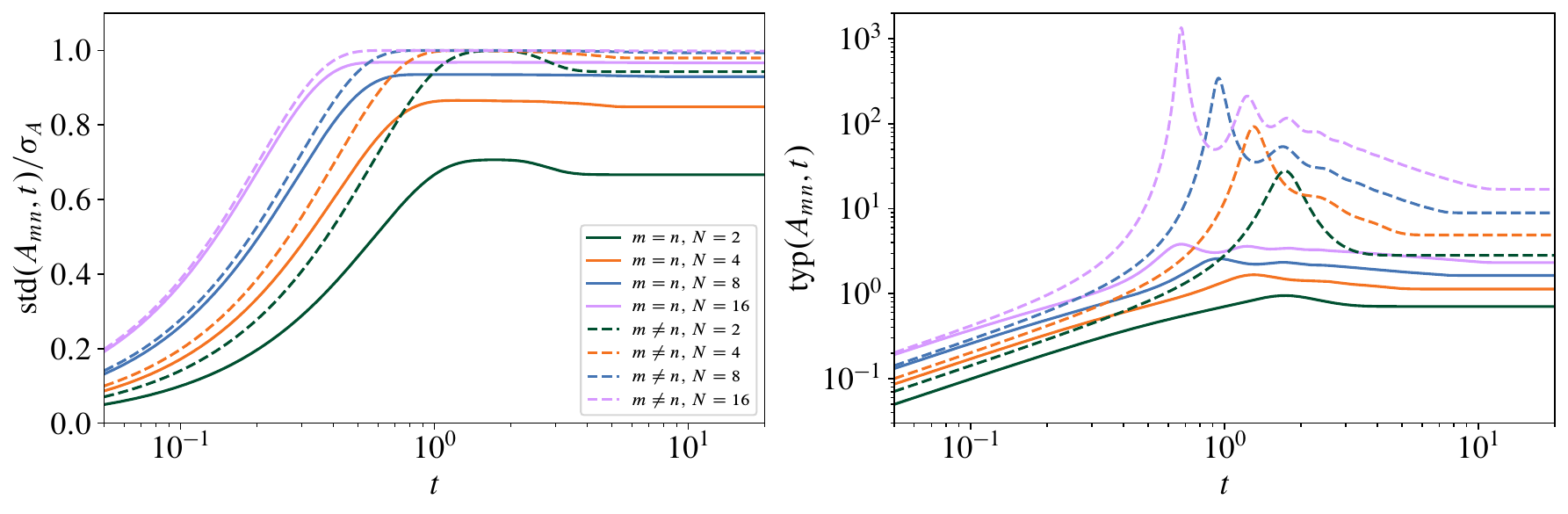}
    \caption{\textbf{Left}: Standard deviation of matrix elements $A_{mn}(t)$ of $A_G(t)$~\eqref{eq:noisyOp}, for an initial operator $A$ sampled from the GUE with width $\sigma_A$. The standard deviation is the square root of the variance~\eqref{eq:varMatElGeneralN}, and is plotted in units of $\sigma_A$ for various $N = 2^k$; solid lines denote diagonal elements ($m=n)$ and dashed lines denote off-diagonal elements ($m \neq n$). \textbf{Right}: Plot of the typicality~\eqref{eq:qubit typ def} of diagonal (solid) and off-diagonal (dashed) matrix elements $A_{mn}(t)$. The typicality is the standard deviation divided by the square root of \eqref{eq:avg square GUE avg}.
    }
    \label{fig:VarFigure}
\end{figure*}

To compute any such variance, we must first calculate the twofold GUE channel~\eqref{eq:nfold channel}, a special case of the quantity
\begin{align}
\label{eq:Var Chan form}
    \varphi^{(2)}_{\mathcal{E}} (A,B;t) \coloneq \expec_{\mathcal{E}} \left[ A_G (t) \otimes B_G(t) \right] \, ,~~
\end{align}
which acts on two copies of $\Hilbert = \Comps^N$. We note that any operator on $\Hilbert^{\otimes 2}$ can be expressed as a linear combination of ``basis'' operators that factorize over the two copies of $\Hilbert$~\cite{MIPT}.

The full expression for the twofold channel~\eqref{eq:Var Chan form}---even in the case $B=A$ relevant to variances---is too unwieldy for the present discussion, and instead appears in App.~\ref{app:k=2 avg}, along with a detailed derivation. As with the derivation of the onefold channel~\eqref{eq:Avg Noisy Chan two terms} in App.~\ref{app:k=1 avg}, we first average the unitaries $V$~\eqref{eq:diagonalizing unitary VG} over the Haar measure~\cite{BnB, Collins_2003}, which involves the \emph{fourfold} Haar channel and its 576 terms. In the case where $B=A$, these terms can be simplified into twelve distinct operators on $\Hilbert^{\otimes 2}$ with time-dependent coefficients,
\begin{align}
    \varphi^{(2)}_{\mathcal{E}} \left( A, t \right) &= \expec_{\mathcal{E}} \left[ \left( \ee^{\ii G t} \hspace{0.2mm} A \hspace{0.2mm}  \ee^{-\ii G t} \right)^{\otimes 2} \right] 
    = \sum\limits_{i=1}^{12} c_i (t) \, \mathcal{O}_i \, ,~~
    \label{eq:twofold 12 ops}
\end{align}
which holds for any random-matrix ensemble $\mathcal{E}$ that is invariant under rotations in $\mathsf{R}=\Unitary{N}$, as described in Sec.~\ref{subsec:eigen pdf}.

The operators $\mathcal{O}_i$ are given by $\ident^{\otimes 2}$, $A^{\otimes 2}$, $\tr(A) \ident \otimes A$, $\tr (A) A \otimes \ident$, $\ident \otimes A^2$, $A^2 \otimes \ident$, along with the foregoing six operators multiplied from the right by the SWAP operator $\mathcal{S}$ on $\Hilbert^{\otimes 2}$, which for any $\ket{a},\ket{b} \in \Hilbert$ acts as $\mathcal{S} \ket{a} \otimes \ket{b} = \ket{b} \otimes \ket{a}$. This structure is ensured by the Haar averages over $V$~\eqref{eq:diagonalizing unitary VG} alone.

The time-dependent coefficients $c_i(t)$, on the other hand, are determined by the spectral PDF $\rho (\bvec{\lambda})$~\eqref{eq:spectral pdf}. These coefficients depend on combinatoric factors determined by Haar averaging, the two-point SFF $\mathcal{R}_2(t)$~\eqref{eq:SFF main}, as well as two higher-point SFFs~\cite{YoshidaCBD, YoshidaSFF}. Most straightforward is the four-point SFF,
\begin{align}
    \label{eq:SFF4 main}
    \mathcal{R}_4 (t) =  \expec_{\mathcal{E}} \sum_{k,\ell, m,n} \ee^{\ii(\lambda_k+\lambda_\ell-\lambda_m-\lambda_n)t} \, , ~~
\end{align}
which is the Fourier transform of the \emph{four}-point eigenvalue density function. Also relevant to the twofold channel~\eqref{eq:twofold 12 ops} is a three-point SFF, corresponding to
\begin{align}
    \label{eq:SFF3 main}
    \mathcal{R}_{4,1} (t) = \expec_{\mathcal{E}}  \sum_{\ell, m,n} \ee^{\ii(\lambda_\ell+\lambda_m-2\lambda_n)t} \, ,~~
\end{align}
where $\mathcal{R}_{4,1}(t)$ denotes the inclusion of one nontrivial Kronecker delta in the summand that defines the four-point SFF $\mathcal{R}_4 (t)$~\eqref{eq:SFF4 main}~\cite{YoshidaSFF}. We explicitly evaluate both the higher-point SFFs $\mathcal{R}_4(t)$ and $\mathcal{R}_{4,1}(t)$ for the GUE in Sec.~\ref{subsec:SFF 4}. The full expression for the generic twofold channel~\eqref{eq:Var Chan form} in terms of various SFFs and operators on $\Hilbert^{\otimes 2}$~\eqref{eq:twofold general} appears in App.~\ref{app:k=2 avg}, though it is not particularly enlightening to behold.

To recover a more straightforward diagnostic of the variance of the average channel~\eqref{eq:Avg Noisy Chan two terms}---and thereby, its typicality---we now consider the variance of individual matrix elements of $A_G(t)$~\eqref{eq:noisyOp}, following the treatment of qubits~\eqref{eq:VarAmn def qubit} in Sec.~\ref{sec:qubits}. 

However, we note that the variance of matrix elements depends explicitly on the initial observable $A$; to avoid bias and to simplify presentation, we sample $A$ itself from the GUE with variance $\sigma_A^2$, in which case we find that
\begin{equation}\label{eq:varMatElGeneralN}
    \expval{\var(A_{mn},t)}^{\vps}_{A} = \sigma_A^2 \left(1-\frac{\kron{m,n}}{N}  \right )  f(t) \left[ 2 - f(t) \right] \, ,~~
\end{equation}
where each factor above is positive semidefinite. 

We plot the standard deviation of matrix elements of $A_G(t)$~\eqref{eq:noisyOp} in the left panel of Fig.~\ref{fig:VarFigure}, for several values of $N=2^k$. For convenience, we sample $A$ from the $N \times  N$ GUE with variance $\sigma_A^2$. As in Fig.~\ref{fig:qubit-var}, we plot the standard deviation of both diagonal elements (solid) and off-diagonal elements (dashed) of $A_G(t)$~\eqref{eq:noisyOp}. We note that the standard deviation is identically zero only at $t=0$, generically nonomonotonic in $t$, and always less than $\sigma_A$.  We also plot the ``typicality'' diagnostic~\eqref{eq:qubit typ def} in the right panel of Fig.~\ref{fig:VarFigure} for several values of $N=2^k$. The typicality is given by the standard deviation divided by the square root of
\begin{align} \label{eq:avg square GUE avg}
    \expval{\operatorname{avg}(A_{mn},t)^2}^{\vps}_{A} &= \sigma_A^2 \left[ \left[ 1 - f(t)\right]^2 + \frac{\kron{m,n}}{N} f(t) \right] \, ,
\end{align}
which follows from the definition of the typicality~\eqref{eq:qubit typ def}. The result is depicted in the right panel of Fig.~\ref{fig:VarFigure}.  Importantly,  Fig.~\ref{fig:VarFigure} reveals that the average~\eqref{eq:Avg Noisy Chan two terms} is representative of a \emph{typical} instance of a noisy unitary~\eqref{eq:noisyOp} for the weak-noise regime $t \ll 1$.  We also observe that diagonal elements of $A_G(t)$~\eqref{eq:noisyOp} appear to be more typical than off-diagonal elements.

\subsection{Higher cumulants}
\label{subsec:higher cum form}

Although we do not evaluate them explicitly herein, one can also use the same RMT methods we use to calculate the average and variance of the noisy channel~\eqref{eq:noisyOp} to compute higher cumulants. The $k$th cumulant extends, e.g., $\var (A,t)$~\eqref{eq:Var Chan form} to $k>2$, and involves the generalized $k$-fold channel~\eqref{eq:nfold channel}
\begin{align}
\label{eq:k-cum chan form}
   \varphi^{(k)}_{\mathcal{E}} (A_1,\dots,A_k;t) = \expec_{\mathcal{E}} \, \left[ \, \bigotimes\limits_{i=1}^{k} \ee^{\ii G t} \hspace{0.2mm} A_i\hspace{0.2mm}  \ee^{-\ii G t} \right]\, , ~
\end{align}
along with various $p$-fold channels for $p<k$.

The evaluation of these $k$-fold GUE channels proceeds in analogy to the discussions in Secs.~\ref{subsec:average chan form} and \ref{subsec:variance chan form}. The first step is to perform the Haar average over the diagonalizing unitaries $V$~\eqref{eq:diagonalizing unitary VG}. The $k$-fold GUE channel involves the $2k$-fold Haar average~\eqref{eq:2k-fold Haar}, as described in App.~\ref{app:k=2 avg}. This average results in $(2k)!^2$ terms, corresponding to $(2k)!/ k!$ distinct operators on on $\Hilbert^{\otimes k}$. As described in  App.~\ref{app:k=2 avg}, the Haar averaging contracts the indices of the various operators $A_i$ in \eqref{eq:k-cum chan form} multiplied by various Weingarten functions~\cite{Samuel1980, Collins_2003, Collins_2006, BnB} and, separately, contractions of factors $\ee^{\pm \ii t \lambda_n}$~\eqref{eq:G spec decomp}. Each term is multiplied by up~to $2k$ factors of the quantity
\begin{equation}
\label{eq:Z(t) def}
    Z(t) \coloneq \tr ( \ee^{- \ii G t } ) = \sum_{m=1}^{N} \ee^{-\ii t \lambda_m}\, ,~
\end{equation}
which contribute to the $2k$-point SFF $\mathcal{R}_{2k}(t)$, along with various lower-point SFFs---e.g., $\mathcal{R}_2(t)$~\eqref{eq:SFF main} and $\mathcal{R}_{4,1}(t)$~\eqref{eq:SFF3 main} both appear in the twofold ($k=2$) channel~\eqref{eq:Var Chan Def}. These lower-point SFFs result from the pairing of eigenvalues due to the Kronecker deltas imposed by the Haar average over the unitaries $V_G$ that diagonalize $G$. All of these SFFs are even functions of $t$, and depend on $\rho (\bvec{\lambda})$ alone (and not the rotation group $\mathsf{R})$. The structure of operators and coefficients that appear in the $k$-fold GUE channel~\eqref{eq:nfold channel} inherit entirely from the fact that $\mathsf{R}=\Unitary{N}$ and that $p(V_G)$ is the Haar measure on $\Unitary{N}$.

Owing to the unwieldy number of terms that appear in the $k$-fold GUE channel~\eqref{eq:nfold channel}, it is best to evaluate such higher-fold channels using software capable of symbolic manipulation. To facilitate such calculations, we have supplied a \texttt{Mathematica} notebook containing functions for deriving all operators present in the $k$-fold GUE channel~\eqref{eq:k-cum chan form} and their time-dependent coefficients. The notebook also details the conversion of these time-dependent coefficients---which involve powers of $Z(t)$~\eqref{eq:Z(t) def}---into generalized SFFs $\mathcal{R}_{2p,\bvec{q}}(t)$ involving $p$ copies of both $Z(t)$~\eqref{eq:Z(t) def} and its conjugate. The $k$-fold GUE channel~\eqref{eq:nfold channel} generally involves the SFFs $\mathcal{R}_{2p,\bvec{q}}(t)$ for $p \leq k$, where $\bvec{q}$ encodes which of the $2p$ eigenvalues involved are constrained to be the same (due to Haar averaging $V_G$). We evaluate these SFFs for the particular case of the GUE appear in Sec.~\ref{sec:SFF}; further details of $k$-fold channels appear in App.~\ref{app:k=2 avg}.

\section{Spectral form factors}
\label{sec:SFF}

One of the fingerprints of quantum chaos---the quantum dynamics that lead to ergodicity and equilibration---is \emph{level repulsion} in the spectrum of the evolution operator $U(t)=\ee^{-\ii G t}$~\cite{YoshidaCBD, YoshidaSFF, BohigasChaos, RMT_SFF, ShenkerRMT, U1FRUC, ConstrainedRUC, MIPT, CDLC1}. This repulsion is an artifact of the \emph{spectral rigidity} of generic chaotic systems \cite{BohigasChaos, CDLC1, RMT_SFF, YoshidaCBD, YoshidaSFF, ShenkerRMT, AmitQuantErgo}. The simplest probe of spectral rigidity is afforded by the ``$r$ ratio''~\cite{rratio}: the average ratio between successive level spacings of $G$. A more detailed probe of spectral statistics is afforded by, e.g., the two-point SFF $\mathcal{R}_2(t)$ \eqref{eq:SFF main}, which is the Fourier transform of the two-point eigenvalue density (or correlation) function of $G$. In particular, the SFF's linear ramp regime---corresponding to $\mathcal{R}_2(t) \sim t$ for $t > \tau_{\text{th}}$, the ``Thouless time''---signals the onset of thermalization. Higher moments of eigenvalue distribution of $G$ are probed via higher-point SFFs $\mathcal{R}_{2k}(t)$, along with more general SFFs. In fact, SFFs for Gaussian Hamiltonians $G$ also appear in the context of black holes \cite{ShenkerRMT, YoshidaCBD}, as well as the various moments of the noisy channel $A_G(t)$ \eqref{eq:noisyOp} discussed in Sec.~\ref{sec:channel def}.

In addition to their role in evaluating higher moments of $A_G(t)$ \eqref{eq:noisyOp}, the various $p$-point SFFs for Gaussian Hamiltonians $G$ are also interesting in their own right. Here, we provide exact analytic expressions for these SFFs with respect to the GUE with arbitrary $N$. In particular, we evaluate the standard two-point SFF $\mathcal{R}_2 (t)$ \eqref{eq:SFF main} in Sec.~\ref{subsec:SFF 2}, the three- and four-point SFFs $\mathcal{R}_{4,1}(t)$ \eqref{eq:SFF3 main} and $\mathcal{R}_{4}(t)$ \eqref{eq:SFF4 main}, respectively, in Sec.~\ref{subsec:SFF 4}, and discuss the generalization to higher-point SFFs in Sec.~\ref{subsec:SFF gen}.

\subsection{Two-point SFF}
\label{subsec:SFF 2}

We first consider the standard, two-point SFF $\mathcal{R}_2(t)$~\eqref{eq:SFF main} that appears in the average GUE channel~$\Avg(A,t)$~\eqref{eq:Avg Noisy Chan two terms},
\begin{align}
    \mathcal{R}_2(t) &\coloneq  \expec_\text{GUE} \sum_{m,n=1}^{N} \ee^{\ii (\lambda_m-\lambda_n) t} \notag \\
    & \; = N + \expec_\text{GUE} \sum_{m \neq n}
    \ee^{\ii (\lambda_m-\lambda_n) t} \, ,~~\label{eq:SFF nonrepeating}
\end{align}
where we separate out the diagonal piece for convenience, and the average is taken with respect to $\rho (\bvec{\lambda})$~\eqref{eq:spectral pdf}. Writing the integral over the spectrum of $G \in \text{GUE}$ explicitly, we define
\begin{align}
    \kappa_2 (t) &\coloneq \frac{1}{N(N-1)} \expec_\text{GUE} \sum_{m \neq n} \ee^{\ii t (\lambda_m-\lambda_n) } \notag \\
    &\;=\int \hspace{-0.7mm} \thed \lambda_1 \thed \lambda_2 \,\rho^{(2)} (\lambda_1, \lambda_2)  \ \ee^{\ii (\lambda_1-\lambda_2) t}  \, ,~
    \label{eq:expIntegral}
\end{align}
where $\rho^{(2)} ( \lambda_1 , \lambda_2)$ is the two-point eigenvalue density function---i.e., the marginal distribution of two eigenvalues, which recovers from integrating $\rho(\bvec{\lambda})$~\eqref{eq:spectral pdf} over $N-2$ of its arguments. The integral~\eqref{eq:expIntegral} vanishes after the Heisenberg time $\tau_{\text{Heis}} = N$, consistent with the Riemann-Lebesgue lemma and the expected behavior of the SFF~\cite{YoshidaSFF, BohigasChaos, RMT_SFF, ShenkerRMT, U1FRUC, ConstrainedRUC, MIPT, CDLC1}. 

The two-point eigenvalue density function can be written
\begin{align}
    \hspace{-0.5mm} \rho^{(2)} (\lambda_1, \lambda_2) &= \hspace{-0.4mm}  \int \hspace{-1mm} \thed \lambda_3 \cdots \thed \lambda_{N} \, \rho (\bvec{\lambda} )  = 
    \frac{ \det_{i,j=1,2} K_N(\lambda_i, \lambda_j) }{N(N-1)}
    \, , ~\label{eq:2eig pdf}
\end{align}
where we used Dyson's theorem~\cite{mehta2004random} in the second line, and have implicitly defined the GUE kernel,
\begin{align}
\label{eq:GUE kernel}
    K_N(\lambda_i,\lambda_j) \coloneq \sum\limits_{n=0}^{N-1} \phi_{n}(\lambda_i) \phi_{n}(\lambda_j) \, ,~~
\end{align}
in terms of the harmonic-oscillator eigenfunctions
\begin{align}
    \label{eq:SHO functions}
    \phi_n (x) = \inprod{x}{n} \coloneq (\sqrt{\pi}2^n n!)^{-1/2}  \ee^{-x^2/2} H_n(x) \, ,~
\end{align}
where $H_n(x)$ is the $n$th (physicist's) Hermite polynomial. The functions $\phi_n(x)$ are the (normalized) eigenfunctions of the harmonic-oscillator Hamiltonian $H= (p^2 + x^2)/2$, with corresponding eigenvalues $E_n=n+1/2$. Details of these conventions---as well as a brief review of orthonormal polynomial methods in RMT---appear in App.~\ref{app:RMT and SFFDetails}.

For convenience, we next rewrite $\kappa_2(t)$~\eqref{eq:expIntegral} as
\begin{equation}
    \label{eq:Kappa2 redef}
    \kappa_2(t) = \frac{\mathcal{R}_{2,d}(t)+\mathcal{R}_{2,c}(t)}{N(N-1)} \, ,~~
\end{equation}
where the subscripts $c$ and $d$ refer to the connected and disconnected contributions, respectively. Defining the ``diagonal'' harmonic-oscillator integrals
\begin{equation}
    \mathcal{I}(t) \coloneq \int \thed \lambda \; \ee^{-\ii t \lambda } \, K_N(\lambda, \lambda) = \sum_{n=0}^{N-1} \matel*{n}{\ee^{-\ii t x}}{n}\, ,~
    \label{eq:I integral}
\end{equation}
along with the ``off-diagonal'' integrals
\begin{equation}
    X_{mn}(t) \coloneq \int \thed \lambda \, \ee^{-\ii t \lambda }\phi_m(\lambda) \phi_n(\lambda) = \matel*{m}{\ee^{-\ii t x}}{n} \,,~
    \label{eq:Xmn integral}
\end{equation}
We can express the disconnected term 
\begin{equation}
    \mathcal{R}_{2,d}(t) = \mathcal{I}(t)\mathcal{I}(-t) = \left| \sum_{m=0}^{N-1} \matel*{m}{\ee^{-\ii t x}}{m} \right|^2 \, ,
    \label{eq:Jd def}
\end{equation}
and for the connected term,
\begin{align}
    \mathcal{R}_{2,c}(t) = &-\sum\limits_{m,n=0}^{N-1} X_{mn}(t)X_{nm}(-t) \notag \\
    = &- \sum_{m,n=0}^{N-1} \matel*{m}{\ee^{\ii t x}}{n}\hspace{-0.4mm} \matel*{n}{\ee^{-\ii t x}}{m} \, ,
    \label{eq:Jc def}
\end{align}
in terms of which we rewrite the SFF~\eqref{eq:SFF main} as
\begin{align}\label{eq:SFF_J_Form}
    \mathcal{R}_2 (t) &= N + \mathcal{R}_{2,d}(t) + \mathcal{R}_{2,c}(t)\,.
\end{align}
The final ingredient required to evaluate the SFF~\eqref{eq:SFF_J_Form} is the form of the matrix elements $\matel*{m}{\ee^{\pm \ii t x}}{n}$, which can be expressed in terms of generalized Laguerre polynomials \cite{schwinger2001quantum, Okuyama_2019} as
\begin{equation}\label{eq:expMatEl}
    \hspace{-2.7mm} \matel*{m}{\ee^{\alpha x}}{n} = \ee^{\frac{\alpha^2}{4}} \hspace{-0.4mm} \begin{cases} \sqrt{\frac{n!}{m!}} 
    \left(\frac{\alpha}{\sqrt{2}}\right)^{m-n}
    L_n^{m-n}\left(-\frac{\alpha^2}{2}\right) & m \geq n \\
    \sqrt{\frac{m!}{n!}} 
    \left(\frac{\alpha}{\sqrt{2}}\right)^{n-m}
    L_m^{n-m} \left(-\frac{\alpha^2}{2}\right) & m \leq n
    \end{cases} \, , 
\end{equation}
so that the SFF~\eqref{eq:SFF main} can be written as
\begin{multline}\label{eq:SFFfinal}
   \hspace{1mm} \mathcal{R}_2(t) = N + 
    \ee^{-\frac{t^2}{2}} \Big[L_{N-1}^{1} \left( \frac{t^2}{2}\right)^2 \\
    - \sum_{m,n=0}^{N-1} (-1)^{m-n}
    L_n^{m-n}\left(\frac{t^2}{2}\right)
    L_m^{n-m}\left(\frac{t^2}{2}\right) \Big]\;,~
\end{multline}
which can be efficiently evaluated for any finite dimension $N$. Further details appear in App.~\ref{app:RMT and SFFDetails}, and we plot the SFF $\mathcal{R}_2(t)$~\eqref{eq:SFFfinal} for various values of $N$ in Fig.~\ref{fig:SFFsFigure}.

\begin{figure*}[t]
    \centering
    \includegraphics[width=\linewidth]{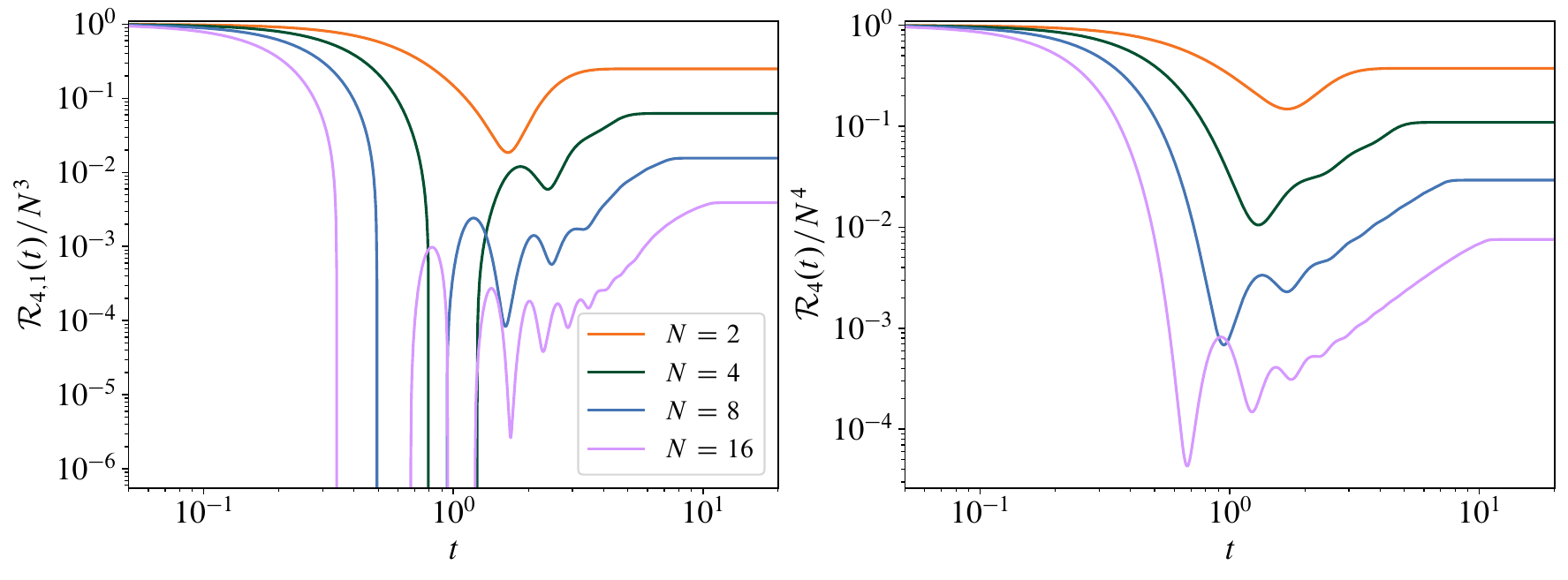}
    \caption{\textbf{Left}: The three-point SFF $\mathcal{R}_{4,1}(t)$~\eqref{eq:SFF3 main} normalized by its $t=0$ value of $N^3$, for $N=2^k$ with $k = 1,2,3,4$. \textbf{Right}: The four-point SFF $\mathcal{R}_4(t)$~\eqref{eq:SFF4 main} normalized by its $t=0$ value of $N^4$, for the same values of $N=2^k$. Both functions bear similarities to $\mathcal{R}_2(t)$~\eqref{eq:SFF main} [see Fig.~\ref{fig:SFFsFigure}].
    }
    \label{fig:SFF34}
\end{figure*}

\subsection{Three-point SFF}
\label{subsec:SFF 3}

We now consider the three--point SFF~\eqref{eq:SFF3 main}, given by
\begin{equation}
    \mathcal{R}_{4,1}(t) = \expec_\text{GUE} \sum_{i,j,k=1}^{N} \ee^{\ii (\lambda_i + \lambda_j - 2 \lambda_k) t } \, ,~~
\end{equation}
which involves the three-point eigenvalue density function
\begin{equation}
    \rho^{(3)}(\lambda_1, \lambda_2, \lambda_3) = \frac{(N-3)!}{N!} \det_{i,j = 1,2,3} K_N (\lambda_i, \lambda_j)\, ,
\end{equation}
in terms of which we write
\begin{align}
    &\hspace{-1mm} \mathcal{R}_{4,1}(t) 
    = \hspace{-0.4mm} \int \hspace{-1mm}  \thed \bvec{\lambda} \, \rho (\bvec{\lambda}) \, 
    \Big[ N + 2 \sum\limits_{i \neq j} 
    \ee^{\ii (\lambda_i - \lambda_j) t} \notag \\ 
    &~~~~~~~~+ \sum\limits_{j\neq k} \ee^{2 \ii (\lambda_i - \lambda_j) t} + \sum\limits_{i \neq j \neq k } \ee^{\ii ( \lambda_i + \lambda_j - 2 \lambda_k ) t} \Big] \, .~ \label{eq:R41 integral}
\end{align}
where the indices $i$, $j$, and $k$ are all distinct in the final term. This expression can be further simplified to
\begin{align}
    \mathcal{R}_{4,1}(t) &= N  + 2 N (N-1) \kappa_2(t)  +  N (N-1) \kappa_2(2t)\notag \\
    &~~~+ N(N-1)(N-2)\kappa_{4,1}(t)\, ,~ \label{eq:R41 kappas}
\end{align}
where $\kappa_2(t)$ is the same function that appears in the two-point SFF $\mathcal{R}_2(t)$~\eqref{eq:expIntegral}, and we have implicitly defined the quantity
\begin{align}\label{eq:K41N>3}
    \kappa_{4,1}(t) \coloneq \hspace{-0.4mm}  \int \hspace{-1mm} \thed \lambda_1 \thed \lambda_2 \thed \lambda_3 \, \rho^{(3)}(\lambda_1,\lambda_2,\lambda_3) \, \ee^{\ii \left( \lambda_1 + \lambda_2 - 2 \lambda_3 \right) t}\, , ~
\end{align}
which we now investigate further. Using the GUE Kernel~\eqref{eq:2eig pdf} and the oscillator eigenfunctions $\phi_n(x)$~\eqref{eq:SHO functions}, we have
\begin{align}\label{eq:K41anyN}
    &\kappa_{4,1}(t) = \frac{(N-3)!}{N!}   \Big[\mathcal{I}(t)^2\mathcal{I}(2t) +2X_{mn}(t)X_{n\ell}(t)X_{\ell m}(-2t)  \notag \\
    &- (2\mathcal{I}(t)X_{mn}(t)X_{nm}(-2t)+\mathcal{I}(2t)X_{mn}(t)X_{nm}(t)) \Big] \, ,~~
\end{align}
with implicit sums on repeated indices. We do not write the full expression for $\mathcal{R}_{4,1}(t)$~\eqref{eq:SFF3 main} in terms of Laguerre polynomials (facilitated through~\eqref{eq:expMatEl}), as it is quite cumbersome and not nearly as useful to behold as the plot of $\mathcal{R}_{4,1}(t)$~\eqref{eq:SFF3 main} that appears in the left panel of Fig.~\ref{fig:SFF34} for several values of $N=2^k$. We remark that the three-point SFF~\eqref{eq:SFF3 main} has similar properties to the two-point SFF~\eqref{eq:SFF main} in Fig.~\ref{fig:SFFsFigure}; however, $\mathcal{R}_{4,1}(t)$~\eqref{eq:SFF3 main} may be negative in the dip regime for $N>2$.

\subsection{Four-point SFF}
\label{subsec:SFF 4}

Next we consider the four-point SFF~\eqref{eq:SFF4 main}, given by
\begin{align}
    \label{eq:SFF4 again}
    \mathcal{R}_4 (t) =  \expec_{\mathcal{E}} \sum_{i,j,k,\ell=1}^N \ee^{\ii(\lambda_i+\lambda_j-\lambda_k-\lambda_\ell)t} \, , ~~
\end{align}
which involves the four-point eigenvalue density function
\begin{equation}
    \rho^{(4)}(\lambda_1, \lambda_2, \lambda_3, \lambda_4) = \frac{(N-4)!}{N!} \det_{i,j = 1,2,3,4} K_N (\lambda_i, \lambda_j)\, .
\end{equation}
As before~\eqref{eq:R41 kappas}, we separate the various terms corresponding to distinct sums over eigenvalues, finding
\begin{align}
    \mathcal{R}_{4}(t) &= N (2N-1) + 4 N (N-1)^2 \kappa_2(t) +  \frac{N!}{(N-2)!} \kappa_2(2t) \notag\\
    &~~+ 2  \frac{N!}{(N-3)!} \kappa_{4,1}(t) +  \frac{N!}{(N-4)!} \kappa_{4}(t) \, ,~ \label{eq:SFF4 kappas}
\end{align}
which involves $\kappa_2(t)$~\eqref{eq:expIntegral}, $\kappa_{4,1}(t)$~\eqref{eq:K41anyN}, and the new quantity
\begin{align}
    \kappa_{4}(t) \coloneq \hspace{-0.4mm}  \int \hspace{-1mm} \thed \vec{\lambda}\, \rho^{(4)}(\vec{\lambda}) \, \ee^{\ii \left( \lambda_1 + \lambda_2 - \lambda_3 - \lambda_4 \right) t}\, , ~
    \label{eq:kappa4}
\end{align}
where we used the shorthand $\vec{\lambda}=\{\lambda_1,\lambda_2,\lambda_3,\lambda_4\}$ for notational convenience. As before, we evaluate $\kappa_{4}(t)$~\eqref{eq:kappa4} using a determinant of the Gaussian kernel $K_N(\lambda_i,\lambda_j)$~\eqref{eq:GUE kernel}, but now with four distinct eigenvalues. We find that
\begin{align}
    \kappa_{4}(t) 
    =
    &\frac{(N-4)!}{N!}[\mathcal{I}^4 - 2\mathcal{I}^2 X^{\vps}_{mn}X^{\vps}_{nm} - 4\mathcal{I}^2 X^{\vps}_{mn}X_{nm}^* \notag \\
    &~~~~ +8 \mathcal{I}X^{\vps}_{mn}X^{\vps}_{n\ell}X_{\ell m}^* 
    +2 X_{mn}^2 X_{nm}^{*2}+X_{mn}^2X_{nm}^2 \notag\\[3pt]
    &~~~~ -4X^{\vps}_{mn} X^{\vps}_{n\ell} X_{\ell k}^* X_{km}^* 
    -2X^{\vps}_{mn} X_{n\ell}^* X^{\vps}_{\ell k} X_{km}^*]
    \, ,~~\label{eq:kappa4 result}
\end{align}
in terms of the harmonic-oscillator integrals $\mathcal{I}(t)$~\eqref{eq:I integral} and $X_{mn}(t)$~\eqref{eq:Xmn integral}, whose time dependence we have suppressed for compactness. Numerical evaluation is again facilitated using~\eqref{eq:expMatEl}, in terms of generalized Laguerre polynomials. Using the above expression for $\kappa_4(t)$~\eqref{eq:kappa4 result}, as well as previously derived expressions for $\kappa_{4,1}(t)$ and $\kappa_2(t)$, we recover an exact analytic expression for the four-point SFF $\mathcal{R}_4(t)$~\eqref{eq:SFF4 kappas} for \emph{any} finite $N$. In turn, this furnishes an exact analytic expression of the variance~\eqref{eq:Var Chan Def} of the GUE channel~\eqref{eq:noisyOp}, which agrees with the single-qubit result~\eqref{eq:VarAmn def qubit} when $N=2$. 

However, the full expression for $\mathcal{R}_4(t)$~\eqref{eq:SFF4 again} is again quite cumbersome and we do not write it explicitly; instead, a plot of $\mathcal{R}_{4}(t)$ appears in Fig.~\ref{fig:SFF34} for several values of $N=2^k$. Its phenomenology mirrors that of the two-point SFF~\eqref{eq:SFF main} in Fig.~\ref{fig:SFFsFigure}.

\subsection{Higher-point SFFs}
\label{subsec:SFF gen}

The methods used to compute the SFFs $\mathcal{R}_2(t)$~\eqref{eq:SFFfinal}, $\mathcal{R}_{4,1}(t)$~\eqref{eq:SFF3 main}, and $\mathcal{R}_{4}(t)$~\eqref{eq:SFF4 again} also extend to arbitrary SFFs of the form $\mathcal{R}_{2p,\bvec{q}}(t)$. The first step is to write $\mathcal{R}_{2p,\bvec{q}}(t)$ as a sum over unique sets of nonrepeating eigenvalues, as for the two-point~\eqref{eq:SFF nonrepeating} and three-point~\eqref{eq:R41 integral} SFFs. This also ameliorates any potential issues that could arise when $2p-q>N$ (so that eigenvalues must be repeated), as with the evaluation of the single-qubit variance channel~\eqref{eq:VarA def qubit}. See also  App.~\ref{app:k=2 avg}.

The result of this expansion is a sum over various functions $\kappa_{2p',\bvec{q'}}(t)$, as with the two-point~\eqref{eq:expIntegral}, three-point~\eqref{eq:R41 kappas}, and four-point~\eqref{eq:SFF4 kappas} SFFs. Each function $\kappa_{2p',\bvec{q'}}(t)$ involves an integral over a complex exponential of $k=2p'-q'$ nonrepeating eigenvalues and the $k$-point eigenvalue density function $\rho^{(k)}(\lambda_1,\dots,\lambda_k)$.  By Dyson's theorem \cite{mehta2004random}, we have that
\begin{equation}
\label{eq:k eigenvalue pdf}
    \rho^{(k)}(\lambda_1, \ldots, \lambda_{k}) = \frac{(N-k)!}{N!} \hspace{-0.4mm} \det_{i,j = 1,\ldots, k} \hspace{-0.7mm} K_N (\lambda_i, \lambda_j)\, , ~
\end{equation}
where the kernel $K_N(\lambda_i,\lambda_j)$~\eqref{eq:GUE kernel} has a simple expression in terms of normalized harmonic-oscillator eigenfunctions~\eqref{eq:SHO functions}.

Hence, $\mathcal{R}_{2p,\bvec{q}}(t)$ can be written as a sum of functions $\kappa_{2p',\bvec{q'}}(t)$ multiplied by $N$-dependent coefficients. The functions $\kappa_{2p',\bvec{q'}}(t)$ are integrals over $\bvec{\lambda}=\{\lambda_1,\dots,\lambda_{2p'-q'}\}$ of various factors $\smash{\ee^{\pm \ii t \lambda_n}}$ times $\smash{\rho^{(k)}(\bvec{\lambda})}$~\eqref{eq:k eigenvalue pdf} for $k=2p'-q'$. Owing to the structure of the kernel $K_N(\lambda_i,\lambda_j)$~\eqref{eq:GUE kernel}, each eigenvalue $\lambda_n$ appears in an expression of the form
\begin{equation}
\label{eq:SFF any k matel}
    \int \hspace{-0.4mm} \thed \lambda \, \ee^{\pm \ii t \lambda} \phi_m (\lambda) \phi_n (\lambda) \ee^{-\lambda^2} = \matel*{m}{\ee^{\pm \ii t x}}{n} \, ,~
\end{equation}
which is simply a matrix element of the complex exponential of the position operator $x$ in the harmonic-oscillator language, which contributes to integral expressions of the form $\mathcal{I}(t)$~\eqref{eq:I integral} or $X_{mn}(t)$~\eqref{eq:Xmn integral}. These functions---and the matrix elements~\eqref{eq:SFF any k matel} themselves---can be written in terms of a sum over finitely many generalized Laguerre polynomials~\eqref{eq:expMatEl}~\cite{schwinger2001quantum, Okuyama_2019}, leading to an exact analytic expression for any $\mathcal{R}_{2p,\bvec{q}}(t)$ eigenvalues.

\section{Outlook}
\label{sec:application}

We have 
developed an exact analytic toolbox for describing noisy quantum evolution that can be smoothly tuned to the identity map. The noise is captured by unitary evolution under a Hamiltonian $G$ drawn from an RMT ensemble that is invariant under rotations $G \to V G V^\dagger$ for $V$ belonging to some group $\mathsf{R}$ (e.g., $\mathsf{R}=\Unitary{N}$ for the GUE). This guarantees that integrals over $G$ simplify greatly to integrals over the \emph{spectrum} of $G$ and over $\mathsf{R}$ with Haar measure. The latter integrals are often known, and the former lead to \emph{spectral form factors}, which are relevant to quantum chaos~\cite{BohigasChaos, RMT_SFF, ShenkerRMT, YoshidaSFF, YoshidaCBD, CDLC1, U1FRUC, ConstrainedRUC, MIPT, AmitQuantErgo, delCampoGUESFF, Okuyama_2019}. The channels may act on the physical system of interest, a subset thereof, and/or on ancilla degrees of freedom representing the environment; various symmetries and constraints may be imposed via projectors (see Refs.~\citenum{U1FRUC, ConstrainedRUC, MIPT} for details).

We compute the average of noisy evolutions $\ee^{\ii G t} \hspace{0.2mm} A \hspace{0.2mm} \ee^{-\ii G t}$~\eqref{eq:noisyOp} with respect to $G$ exactly. The result $\Avg (A,t)$ takes the form of a time-dependent \emph{depolarization} channel~\eqref{eq:Avg Noisy Chan two terms}~\cite{QC_book}. Importantly, the time-dependent amplitudes (\emph{i}) depend on the noise strength $\sigma$ and duration $t$ only through the product $\sigma t$, (\emph{ii}) depend on $t$ only through the two-point SFF $\mathcal{R}_2(\sigma t)$~\eqref{eq:SFF main}, (\emph{iii}) are \emph{nonmonotonic} functions of $\sigma t$, and (\emph{iv}) never saturate the Haar-random limit $\Avg (A,t) \propto \ident$ (see Fig.~\ref{fig:SFFsFigure}). We also compute the variance of matrix elements of $\ee^{\ii G t} \hspace{0.2mm} A \hspace{0.2mm} \ee^{-\ii G t}$~\eqref{eq:noisyOp}, which we use to diagnose typicality~\eqref{eq:qubit typ def}. We observe that the average channel $\Avg (A,t)$~\eqref{eq:Avg Noisy Chan two terms} reflects a typical instance of noisy evolution $\ee^{\ii G t} \hspace{0.2mm} A \hspace{0.2mm} \ee^{-\ii G t}$~\eqref{eq:noisyOp}---in the sense that $\operatorname{typ}(A_{mn},t) \ll 1$---in the weak-noise limit $\sigma t \lesssim 1$. We also sketch the evaluation of higher cumulants to diagnose the noisy evolution in detail, and have included a \texttt{Mathematica} notebook for doing so~\cite{notebook}.

The foregoing calculations apply to noisy evolution generated by any ensemble invariant under $\mathsf{R}=\Unitary{N}$, and depend only on the corresponding SFF $\mathcal{R}_2(t)$~\eqref{eq:SFF main} for the spectral distribution. We compute the SFFs $\mathcal{R}_2(t)$~\eqref{eq:SFF main}, $\mathcal{R}_{4,1}(t)$~\eqref{eq:SFF3 main}, and $\mathcal{R}_4(t)$~\eqref{eq:SFF4 main} for the GUE \emph{exactly} for any finite $N$ for the first time in terms of generalized Laguerre polynomials (see Fig.~\ref{fig:SFF34}). Our results for $\mathcal{R}_2(t)$~\eqref{eq:SFF main} correct typos that appear 
in Ref.~\citenum{delCampoGUESFF}. We also sketch how to evaluate the higher-point SFFs relevant to higher cumulants of the noisy channel, which we also expect to be of interest to the study of quantum chaos and even black holes.

Although we primarily restricted to the GUE for simplicity, the methods we outline extend to other rotation-invariant random-matrix ensembles $\mathcal{E}$. Importantly, we expect that averaging over noisy evolutions results in more interesting effective channels for ensembles beyond the GUE, as is the case for the Gaussian orthogonal (GOE) and symplectic (GSE) ensembles~\cite{GOEfuture}, whose averages do \emph{not} have the form of a depolarizing channel. Our methods are also compatible with the three standard Wishart-Laguerre ensembles~\cite{Wishart, Livan_2018}, and the various Altland-Zirnbauer ensembles~\cite{AZensembles}. In general, the choice of ensemble should be dictated by experimental considerations and symmetry properties. We also note that additional structure (e.g., symmetries and/or constraints) can be enforced on the noisy unitaries using appropriate projectors~\cite{U1FRUC, ConstrainedRUC, MIPT}. We expect the application of this method to other ensembles---and to the diagnosis of quantum phases of matter and the robustness to noise of various protocols for tomography---to be of great utility to various areas of quantum science.

There are several promising applications for our methodology and directions for future work. In the context of condensed matter, our methods allow one to establish the existence of (possibly nonequilibrium) quantum phases of matter by probing the robustness of the universal properties to noise. In the context of atomic, molecular, and optical physics, it is relevant to identifying whether the dominant sources of noise are coherent or incoherent, and to make quantitative predictions for time evolutions subject to coherent noise. Our framework is also extremely relevant to QEC in various contexts~\cite{CalderbankGood, SteaneQEC1, SteaneQEC2, DiVincenzoShorQEC, Kitaev97, GottesmanFTQC, Knill_2000, Dennis_2002, TemmeZNE, Lieu2020}: Our exact results for qubits (without averaging) can be used to test the efficacy of error-correction protocols by modeling all noise as (possibly correlated) single-qubit unitary errors; our results for general $N$ may also be useful to probing the efficacy of error-correction protocols. Furthermore and most importantly, our framework may be used to improve randomized benchmarking~\cite{Emerson05, Levi07, Knill08, Dankert09, Magesan11, Wallman15} either by (\emph{i}) incorporating block structure to diagnose noise that respects a particular symmetry or obeys a particular constraint or (\emph{ii}) drawing $G$ from an ensemble other than the GUE, resulting in a channel that is \emph{not} in the depolarizing form, yielding more detailed information than Haar-random benchmarking (or any 2-design). Finally, our methods may be used to diagnose the sensitivity to noise between shots of various protocols that require multiple copies of a quantum state $\rho$, including cross-entropy diagnostics and all forms of state and process tomography~\cite{HuangShadow, HuangAdvantage, HuangEfficient}.

\emph{Data availability}.---The \texttt{Mathematica} notebook used to perform Haar averaging is available on Zenodo~\cite{notebook}. 
\\
\\

\section*{Acknowledgements}
\label{sec:cheers}

This work was supported  by the Air Force Office of Scientific Research under Award No. FA9550-20-1-0222 (MO, OH, RN), and the US Department of Energy via Grant No. DE-SC0024324 (AJF).

\appendix
\renewcommand{\thesubsection}{\thesection.\arabic{subsection}}
\renewcommand{\thesubsubsection}{\thesubsection.\arabic{subsubsection}}
\renewcommand{\theequation}{\thesection.\arabic{equation}}

\section{Gaussian unitary ensemble}
\label{app:GUE}

For the sake of clarity, we explicitly define the  Gaussian unitary ensemble GUE$(N,\sigma)$ of complex $N \times N$ Hermitian matrices~\cite{mehta2004random}, and discuss various properties of this random-matrix ensemble. We provide two equivalent definitions:
\begin{enumerate}
    \item The ensemble GUE$(N,\sigma)$ consists of all $N \times N$ matrices $G$ whose diagonal elements $G_{ii} \in \Reals$ are drawn from the (real) normal distribution $\mathcal{N}_{\Reals} (0,\sigma^2)$ and whose upper-triangular elements $G_{ij} \in \Comps$ for $i<j$ are drawn from the (complex) normal distribution $\mathcal{N}_{\Comps} (0,\sigma^2)$. The lower-triangular elements are defined as $G_{ij} \coloneq G^*_{ji}$ for $i>j$. 
    \item The ensemble GUE$(N,\sigma)$ consists of all $N \times N$ matrices $G$ of the form $G \coloneq (X+X^{\dagger})/2$, where \emph{all} components of the matrix $X$ are independently drawn from the (complex) normal distribution $\mathcal{N}_{\Comps} (0,\sigma^2/2)$. 
\end{enumerate}
Note that the complex normal distribution satisfies $\mathcal{N}_{\Comps} (0,\sigma) \sim \mathcal{N}_{\Reals} (0,\sigma^2/2) + \ii \, \mathcal{N}_{\Reals} (0,\sigma^2/2)$; i.e., the real and complex parts are independently drawn from a normal distribution with half the variance. As a result, \emph{all} elements $G_{ij}$ of the Gaussian matrix $G \in \text{GUE} (N,\sigma)$ have variance $\sigma^2$. 

Using straightforward algebraic manipulations, the product of the probability density functions (PDFs) for the individual matrix elements of a Gaussian unitary matrix $G \in \text{GUE} (N,\sigma)$ can be rewritten as the following PDF for $G$ itself,
\begin{equation}
\label{eq:P of G GUE}
    P_{\text{GUE}} (G,\sigma) = \frac{1}{Z} \exp \left[ - \frac{1}{2 \sigma^2} \tr ( G^2 ) \right] \, ,~~
\end{equation}
subject to the constraint that $G = G^{\dagger}$ is Hermitian, where the ``partition function'' $Z$ ensures normalization of $P$ \eqref{eq:P of G GUE}. The corresponding one-point eigenvalue density function $\rho_1(\lambda)$ realizes Wigner's semicircle law~\cite{mehta2004random},
\begin{equation}
    \label{eq:GUE semicircle}
    \rho_1 (\lambda) = \frac{1}{2\pi \sigma^2 N} \sqrt{ 4 N \sigma^2 - \lambda^2}  +    \cdots ,
\end{equation}
to leading order as $N \to \infty$; the ``$\cdots$'' term above reflects subleading corrections, which vanish faster than $N^{-1/2}$.

For convenience, we take $\sigma = 1/\sqrt{2}$ throughout the main text, in which case the PDF for a complex, Hermitian matrix $G$ sampled from the GUE takes the form
\begin{equation}
\label{eq:P of G GUE nice}
    P_{\text{GUE}} (G) = \frac{1}{\Omega} \exp \left[ - \tr ( G^2 ) \right] \, ,
\end{equation}
and the ``strength'' of the noisy unitary $U_G = \ee^{-\ii t G}$ is controlled simply by the time $t$. This choice is particularly convenient for the computation of the spectral form factor (related to the average channel) for arbitrary $N$ in Sec.~\ref{sec:SFF}, because the physicists' Hermite polynomials form a basis with respect to the Gaussian measure \eqref{eq:P of G GUE nice}. For this choice of $\sigma = 2^{-1/2}$, the corresponding semicircle law \eqref{eq:GUE semicircle} takes the familiar form
\begin{equation}
    \label{eq:GUE semicircle nice}
    \rho_1 (\lambda) = \frac{1}{\pi N} \sqrt{2 N - \lambda^2} \, , ~~ 
\end{equation}
up to corrections that vanish as $1/N$ or faster.    
\section{Ensemble averaging}
\label{app:averaging}
Here we provide explicit details of the ensemble-averaging procedure summarized in Sec.~\ref{sec:channel def}. We define rotational invariance in App.~\ref{app:rotInv}, showing that the eigenvectors of Gaussian unitaries are Haar random and other details relevant to Sec.~\ref{subsec:eigen pdf}. In App.~\ref{app:k=1 avg}, we utilize known results for Haar integrals to obtain the exact form of the average noisy channel~\eqref{eq:Avg Noisy Chan two terms} that is the main result of this work and Sec.~\ref{subsec:average chan form}. In App.~\ref{app:k=2 avg}, we use similar methods to derive the twofold channel~\eqref{eq:twofold 12 ops}, which we use to diagnose typicality of the average channel in Sec.~\ref{subsec:variance chan form}.
\subsection{Rotation Invariance}
\label{app:rotInv}
A powerful constraint on RMT ensembles derives from \emph{rotation invariance}, which we now elucidate. An RMT ensemble $\mathcal{E}$ corresponds to a PDF $P_{\mathcal{E}}(G)$ over matrices $G$ belonging to a subspace $\mathcal{E}$ of the space $M_N (\Comps)$ of complex $N \times N$ matrices. The subspace $\mathcal{E}$ is generally specified by defining how the components $G_{ij}$ of $G$ are sampled (see for example the construction of elements $G$ of the GUE in App.~\ref{app:GUE}). 

An  random-matrix ensemble $\mathcal{E}$ with PDF $P_{\mathcal{E}}(G)$ is invariant under a \emph{rotation group} $\mathsf{R}$ if the following conditions hold:
\begin{enumerate}
    \item For any $G_0 \in \mathcal{E}$ and for all $Q \in \mathsf{R}$, we have that
    \begin{equation}
    \label{eq:rotation closure def}
        G' = Q \hspace{0.2mm} G_0 \hspace{0.2mm} Q^\dagger \in \mathcal{E} \, .~
    \end{equation}
    \item For any $G_0 \in \mathcal{E}$ and $Q \in \mathsf{R}$, we have that
    \begin{equation}
    \label{eq:rotation invariance def}
        P_{\mathcal{E}}(G') = P_{\mathcal{E}} (Q \hspace{0.2mm} G_0 \hspace{0.2mm} Q^\dagger ) = P_{\mathcal{E}} (G) \, .~
    \end{equation}
    \item For any $G_0 \in \mathcal{E}$, the rotation matrix $V^{\smash \vpd}_{\smash G}$ that diagonalizes $G_0$ is itself an element of $\mathsf{R}$, i.e., 
    \begin{equation}
    \label{eq:rotation diagonalizing}
        V_{\smash G}^{\smash \dagger} \hspace{0.2mm} G_0 \hspace{0.2mm}  V_{\smash G}^{\smash \vpd} = \operatorname{diag}(\bvec{\lambda}) ~ \implies ~ V^{\smash \vpd}_{\smash G} \in \mathsf{R} \, . ~~
    \end{equation}
\end{enumerate}
The first condition ensures that conjugating elements of $\mathcal{E}$ by elements of $\mathsf{R}$ returns an element of $\mathcal{E}$. The second condition is the standard meaning of rotation invariance of $P_{\mathcal{E}}(G)$.  The third condition is essential to simplifying ensemble averages; note that the physically relevant choices of  $\mathcal{E}$ are ensembles of Hermitian matrices of exponentials thereof, which are always diagonalizable. It is sufficient for the above if (\emph{i}) the diagonalized version of every element of $\mathcal{E}$ is also an element of $\mathcal{E}$ and (\emph{ii}) the image of the adjoint action of $\mathsf{R}$ on $\mathcal{E}$ is precisely $\mathcal{E}$. We also comment that, in the presence of block structure~\cite{U1FRUC, ConstrainedRUC, MIPT}, the conditions above hold within each block individually. 

Before using rotation invariance to simplify averages over the RMT ensemble $\mathcal{E}$, we first rewrite the PDF $P_{\mathcal{E}}(G)$ in terms of the \emph{spectral decomposition} of $G \in \mathcal{E}$,
\begin{equation*}
    \tag{\ref{eq:spectral pdf}}
    P_{\mathcal{E}} (G) \, \thed G = \rho (\bvec{\lambda}) \, p (V^{\vpd}_{\smash G})  \hspace{0.2mm}  \thed \bvec{\lambda} \hspace{0.2mm} \thed V^{\vpd}_{\smash G} \, ,~~
\end{equation*}
where $\bvec{\lambda}=\{\lambda_1,\lambda_2,\dots,\lambda_N\}$ contains the $N$ eigenvalues of $G$, $\rho (\bvec{\lambda})$ is the joint PDF of the $N$ eigenvalues of $G$, and $p(V^{\smash \vpd}_{\smash G})$ is the PDF for the unitary matrix $V^{\smash \vpd}_{\smash G}$ that diagonalizes $G$, and whose columns are normalized eigenvectors of $G$.

As a sanity check for the validity of~\eqref{eq:spectral pdf}, we may check that both sides of~\eqref{eq:spectral pdf} contain the same number of degrees of freedom. We begin by writing $G$ in the form
\begin{equation}
    \label{eq:G spec decomp}
    G = \sum\limits_{n=1}^{N}  \lambda_n \, \BKop{\psi_n}{\psi_n} = \sum\limits_{n=1}^{N} \lambda_n \, V^{\vpd}_{\smash G} \, \BKop{n}{n} V^{\dagger}_{\smash G} \, . 
\end{equation}
Additionally, the $n$th column of $V^{\smash \,}_{\smash G}$ gives the components of the eigenvector $\ket{\psi_n}$ in the $\{ \ket{n} \}$ basis, i.e.,
\begin{equation}
\label{eq:diagonalizing unitary VG}
    \matel*{m}{V_{\smash G}}{n} = \inprod{m}{\psi_n} ~ \implies ~ V_{\smash G} = \sum\limits_{n=1}^{N} \, \BKop{\psi_n}{n} \, ,~
\end{equation}
so that $G_{\text{diag}} = V^{\dagger}_{\smash G} \hspace{0.4mm} G \hspace{0.4mm}  V^{\vpd}_{\smash G}$ is diagonal in the $\{ \ket{n} \}$ basis. Importantly, $G_{\text{diag}}$ commutes with \emph{any} diagonal unitary $U_{\text{diag}} = \operatorname{diag}(\ee^{\ii \theta_1}, \dots, \ee^{\ii \theta_N})$---equivalently, each ket $\ket{\psi_n}$ in the outer-product definition of $V^{\,}_{\smash G}$ \eqref{eq:diagonalizing unitary VG} is only unique up to a complex phase $\ee^{\ii \theta_n}$. Consequently, the diagonalizing unitary $V^{\,}_{\smash G}$ \eqref{eq:diagonalizing unitary VG} is only \emph{uniquely} defined up to some $U_{\text{diag}}$.
Hence, the diagonalizing unitary $V^{\,}_{\smash G}$ \eqref{eq:diagonalizing unitary VG} is specified by $\smash{N^2-N}$ real, compact parameters, compared to $\smash{N^2}$ for an arbitrary element of $\Unitary{N}$. Combined with the $N$ eigenvalues $\{ \lambda_n\}$ leads to $\smash{N^2}$ real parameters, as needed to specify an element of the GUE~\cite{mehta2004random, Livan_2018, eynard2018random}.

The change of variables \eqref{eq:spectral pdf} is useful because all conjugation-invariant functions of a matrix $G$---e.g., the probability distribution $P_{\mathcal{E}} (G)$---can be expressed in terms of traces of powers of $G$, which can be expressed entirely in terms of the eigenvalues of $G$ (since the determinant is related to the traces via the Cayley-Hamilton Theorem). For Gaussian ensembles, e.g., only $\smash{\trace (G^2)}$ and $\trace (G)$ are needed \cite{PRGaussianEnsemble}. As derived in, e.g., Ref.~\citenum{mehta2004random}, for the GUE with PDF $P_{\mathcal{E}}(G)$ \eqref{eq:P of G GUE}, we have that
\begin{equation}
\label{eq:p eigs GUE}
    \rho ( \bvec{\lambda} ) = \frac{1}{\Omega}  \ee^{- \bvec{\lambda}^2 / 2 \sigma^2} \, \abs{ \Delta (\bvec{\lambda} ) }^2 \, , ~~
\end{equation}
where $\bvec{\lambda}^2 = \lambda_1^2  + \cdots + \lambda_N^2$ and the object $\Delta (\bvec{\lambda})$ is simply the \emph{Vandermonde determinant}, given by
\begin{equation}
    \label{eq:Vandermonde det}
    \Delta (\bvec{\lambda}) \coloneq \prod\limits_{i < j} ( \lambda_i - \lambda_j ) \, ,~~
\end{equation}
which comes from the change of variables captured by \eqref{eq:spectral pdf}. Other choices of ensembles $\mathcal{E}$ will result in different joint eigenvalue density functions $\rho (\bvec{\lambda})$ above.

We now invoke rotation invariance of $\mathcal{E}$, and specifically, of $P_{\mathcal{E}}(G)$ \eqref{eq:rotation invariance def}. The requirement that the diagonalizing matrix $V_G$ is an element of $\mathsf{R}$ \eqref{eq:rotation diagonalizing} immediately implies that
\begin{equation}
   \label{eq:diagP}
    P_{\mathcal{E}} (G) = P_{\mathcal{E}} (V_{\smash G}^{\smash \dagger} \hspace{0.2mm} G \hspace{0.2mm} V^{\smash \vpd}_{\smash G}) = P_{\mathcal{E}} (D)\,, ~~
\end{equation}
where $D = V_{\smash G}^{\smash \dagger} \hspace{0.2mm} G \hspace{0.2mm} V^{\smash \vpd}_{\smash G} = \operatorname{diag}(\bvec{\lambda})$ is the diagonalized version of $G \in \mathcal{E}$, so schematically, $P_{\mathcal{E}} (D) \sim \rho (\bvec{\lambda})$. Importantly, since $\mathsf{R}$ is a group, rotation invariance of $P_{\mathcal{E}}(G)$ \eqref{eq:rotation invariance def} also implies that
\begin{equation}
\label{app:diagP2}
    P_{\mathcal{E}} (G)  = P_\mathcal{E}(W^\dagger V_{\smash G}^{\smash \dagger}  U^\dagger \hspace{0.2mm} G \hspace{0.2mm}  U V^{\smash \vpd}_{\smash G} W ) = P_{\mathcal{E}} (D) \, ,  ~~
\end{equation}
for any $U,W \in \mathsf{R}$. Combining the foregoing relation with the spectral decomposition \eqref{eq:spectral pdf}, we conclude that
\begin{equation}
    \label{eq:p VG Haar}
    p (U \hspace{0.2mm} V^{\smash \vpd}_{\smash G} \hspace{0.2mm} W) = p (V^{\smash \vpd}_{\smash G}) ~~ \forall \, U,W \in \mathsf{R} \, ,~~
\end{equation}
which is known as \emph{left-right invariance} of $p (V^{\smash \vpd}_{\smash G})$ over $\mathsf{R}$. This is ensured by the requirement that $V^{\smash \vpd}_{\smash G} \in \mathsf{R}$ \eqref{eq:rotation diagonalizing}. Importantly, the only measure $p (V^{\smash \vpd}_{\smash G})$ satisfying left-right invariance \eqref{eq:p VG Haar} for a compact matrix group $\mathsf{R}$ is the \emph{Haar} measure, which is uniform over $\mathsf{R}$~\cite{Simon1995RepresentationsOF, Collins_2003, Collins_2006, Collins_2009, collins2021weingarten}. This is a useful result because averages over the Haar measure $p (V^{\smash \vpd}_{\smash G})$ are generally the most straightforward of all RMT ensembles, and have been worked out for the compact matrix groups $\mathsf{R}$ of interest~\cite{BnB, YoshidaCBD,  Simon1995RepresentationsOF, Collins_2003, Collins_2006, Collins_2009, collins2021weingarten, MatsumotoCOE_2012,Matsumoto_AZ_Wg_2013}.

\subsection{Onefold average}
\label{app:k=1 avg}

The most important quantity we compute is the onefold channel~\eqref{eq:Avg Chan Def}, associated with the \emph{average} over noisy unitary evolutions~\eqref{eq:noisyOp} generated by GUE Hamiltonians; here, we provide a detailed derivation of the result~\eqref{eq:Avg Noisy Chan two terms}. To start, it is useful to write the noisy unitary in the form
\begin{align}
    U_G (t) &\coloneq \ee^{-\ii G t} = \sum_{m=1}^{N} \ee^{-\ii \lambda_m t} \, V^{\smash \vpd}_{\smash G} \BKop{m}{m} V^{\smash \dagger}_{\smash G} \notag \\
    &= \sum_{i,j,m=1}^{N} \ee^{-\ii \lambda_m t} V^{\smash \vps}_{im} V^{\smash *}_{mj} \, \BKop{i}{j} \, ,~~
    \label{appeq:timeEvolExpanded}
\end{align}
in the computational basis in which the components of $G$ are drawn, where $V$ diagonalizes $G$ \eqref{eq:rotation diagonalizing}, so that the $m$th column of $V$ is a normalized eigenvector of $G$ with eigenvalue $\lambda_m$---i.e., $V_{im} 
= \inprod{i}{\psi_m}$, where $G \ket{\psi_m} = \lambda_m \ket{\psi_m}$.

We then consider the Heisenberg evolution of an observable $A$ under $U_G(t)$~\eqref{appeq:timeEvolExpanded}, captured by the adjoint action
\begin{equation*}
\tag{\ref{eq:noisyOp}}
    A(t) = U_{\smash G}^{\smash \dagger} \hspace{0.2mm} A \hspace{0.2mm} U_{\smash G}^{\smash \vpd} = \ee^{\ii Gt} \hspace{0.2mm} A \hspace{0.2mm} \ee^{-\ii Gt}\, ,~~
\end{equation*}
where the evolution of a density matrix $\rho$ under $U_G(t)$~\eqref{appeq:timeEvolExpanded} is captured by taking $A \to \rho$ and $t \to -t$ above. Expanding $A(t)$~\eqref{eq:noisyOp} in the computational basis, we have
\begin{align}
    A(t) &= \hspace{-1.5mm} \sum\limits_{i,j,m,n,\mu,\nu} \hspace{-1.5mm} \ee^{\ii ( \lambda_m - \lambda_n) t} V^{\smash \vps}_{\mu m} V^{\smash *}_{m i} A^{\smash \vps}_{\smash ij} V^{\smash \vps}_{j n} V^{\smash *}_{n \nu} \BKop{\mu}{\nu} \, ,~\label{eq:onefold in basis}
\end{align}
which we then simplify by averaging $V$ with respect to $p(V)$~\eqref{eq:p VG Haar}. Owing to rotation invariance, as discussed in App.~\ref{app:rotInv}, averaging $V$ over $\Unitary{N}$ is straightforward, because $p(V)$ is simply the (uniform) Haar measure. 

The average over $V$ in $A(t)$~\eqref{eq:onefold in basis} corresponds to the twofold Haar average over $\Unitary{N}$~\cite{BnB, Simon1995RepresentationsOF, Collins_2003, Collins_2006, Collins_2009, collins2021weingarten}, i.e.,
\begin{align}
    &~\int \thed \mu (V) \, V^{\smash \vps}_{\smash a_1 b_1} V^{\smash \vps}_{\smash a_2 b_2} V^{\smash *}_{\smash a_3 b_3} V^{\smash *}_{\smash a_4 b_4} = \frac{1}{N^2 -1} \times \notag \\
    &\hspace{-1mm} \left[ \delta^{a_1 b_1 a_2 b_2}_{a_3 b_3 a_4 b_4} + \delta^{a_1 b_1 a_2 b_2}_{a_4 b_4 a_3 b_3} - \frac{1}{N} \left( \delta^{a_1 b_1 a_2 b_2}_{a_3 b_4 a_4 b_3}  + \delta^{a_1 b_1 a_2 b_2}_{a_4 b_3 a_3 b_4} \right) \right] ,  \label{eq:twofold Haar avg}
\end{align}
where, for convenience, we have used the shorthand
\begin{equation}
    \delta^{\, ab\dots}_{\alpha \beta \dots} = \kron{a,\alpha} \kron{b,\beta} \cdots \,, ~~
\end{equation}
and applying the result~\eqref{eq:twofold Haar avg} to $A(t)$~\eqref{eq:onefold in basis} leads to
\begin{align*}
    &\int \thed \mu (V) \, A(t) = \sum_{\substack{i,j,m \\ n , \mu,\nu}} \ee^{\ii ( \lambda_m - \lambda_n ) t} \frac{1}{N^2-1}  A^{\smash \vps}_{\smash ij} \BKop{\mu}{\nu} \, \times    \\
    \hspace{-2mm} &\left[ \kron{\mu,i} \kron{j,\nu} + \kron{\mu,\nu} \kron{m,n} \kron{i,j}  - \frac{1}{N}  \left( \kron{\mu,i} \kron{m,n} \kron{j,\nu} + \kron{\mu,\nu} \kron{i,j} \right) \right] \, , 
\end{align*}
and summing over $i$ and $j$ leads to 
\begin{align}
    &\int \thed \mu (V) \, A(t) = \sum_{m,n,\mu,\nu} \frac{\ee^{\ii (\lambda_m - \lambda_n )t} }{N^2-1} \BKop{\mu}{\nu} \times \notag \\
    \hspace{-2mm} &\left[ \left( 1- \frac{\kron{m,n}}{N} \right) A_{\mu,\nu} +\tr (A) \left( \kron{m,n}- \frac{1}{N} \right) \kron{\mu,\nu} \right] \, , ~ \notag 
\end{align}
and, finally, summing over $\mu$ and $\nu$ gives
\begin{equation*}
    = \sum\limits_{m,n=0}^{N-1} \frac{\ee^{\ii (\lambda_m - \lambda_n )t} }{N^2-1} \left[ \left( 1 - \frac{\kron{m,n}}{N} \right) A + \tr (A) \left( \kron{m,n} - \frac{1}{N} \right) \ident \right] \, ,~ \label{eq:onefold avg unitary done}
\end{equation*}
and for further simplification, we recall the definition
\begin{equation*}
\tag{\ref{eq:Z(t) def}}
    Z(t) \coloneq \tr ( \ee^{- \ii G t } ) = \sum_{m=1}^{N} \ee^{-\ii t \lambda_m}\, ,
\end{equation*}
in terms of which we can write
\begin{subequations}
\label{eq:onefold Z relations}
\begin{align}
    \sum_{m,n=1}^{N} \ee^{\ii (\lambda_m-\lambda_n)t} \left(1-\frac{\kron{m,n}}{N}\right) &= \abs{Z(t)}^2 - 1  \\
    \sum_{m,n=1}^{N} \ee^{\ii (\lambda_m -\lambda_n) t} \left( \kron{m,n}-\frac{1}{N}\right) &= N - \abs{Z(t)}^2 \, , ~~
\end{align}
\end{subequations}
so that we can write the Haar average of $A(t)$~\eqref{eq:onefold in basis} as
\begin{align}
    \label{appeq:Haar Averaged Channel}
    \hspace{-2.5mm} \int\hspace{-1mm} \thed \mu(V) A(t) = \frac{\abs{Z(t)}^2-1}{N^2-1} A + \frac{N^2-\abs{Z(t)}^2}{N^2-1}\frac{\trace(A)}{N} \ident \, , \hspace{-2mm}
\end{align}
and noting that the quantity
\begin{equation}\label{eq:SFF from Z}
    \mathcal{R}_2(t) = \int_{\mathcal{E}}  \thed \bvec{\lambda} \, \rho(\bvec{\lambda}) \abs{Z(t)}^2 \, ,~~
\end{equation}
is the two-point SFF~\eqref{eq:SFF main}~\cite{YoshidaCBD, RMT_SFF, YoshidaSFF, CDLC1,  U1FRUC, ConstrainedRUC, MIPT, ShenkerRMT}, averaging \eqref{appeq:Haar Averaged Channel} over the eigenvalues $\bvec{\lambda}$ with respect to $\rho (\bvec{\lambda})$~\eqref{eq:spectral pdf} leads to
\begin{equation}\label{appeq:Averaged Channel}
    \varphi^{(1)}_{\mathcal{E}} (A,t) = \frac{\mathcal{R}_2(t)-1}{N^2-1}A + \frac{N^2-\mathcal{R}_2(t)}{N^2-1}\frac{\trace(A)}{N} \ident \, ,
\end{equation}
which recovers the result~\eqref{eq:Avg Noisy Chan two terms} of the main text upon identifying $f(t)$~\eqref{eq:f(t) def} in terms of the SFF~\eqref{eq:SFF from Z}. Note that this particular form~\eqref{appeq:Averaged Channel} holds for \emph{any} random-matrix ensemble that is invariant under rotations in $\mathsf{R}=\Unitary{N}$, as described in App.~\ref{app:rotInv}, where $\mathcal{R}_2(t)$~\eqref{eq:SFF from Z} is determined by $\rho(\bvec{\lambda})$~\eqref{eq:spectral pdf} alone.

\subsection{Twofold and higher averages}
\label{app:k=2 avg}

Diagnosing the typicality of the average channel~\eqref{eq:Avg Noisy Chan two terms} involves calculating the variance of matrix elements of $A(t)$~\eqref{eq:onefold in basis}. That, in turn,  requires the evaluation of the twofold GUE channel, whose most general form is given by
\begin{align*}
    \varphi^{(2)}_{\mathcal{E}} (A,B;t) &\coloneq \expec_{\mathcal{E}} \left[ \left( \ee^{\ii G t} A \ee^{-\ii G t} \right) \otimes \left( \ee^{\ii G t} B \ee^{-\ii G t} \right)  \right] \\
    &\; = \int \thed G \, P_{\mathcal{E}}(G) \, A_G(t) \otimes B_G(t)\, ,~\tag{\ref{eq:Var Chan form}}
\end{align*}
which acts on $\Hilbert^{\otimes 2} = \Comps^N \otimes \Comps^N$. Note that generic operators on $\Hilbert^{\otimes 2}$ may be decomposed onto a basis of operators~\cite{MIPT} that factorize over the two copies of $\Hilbert$. The case relevant to calculating variances corresponds to $B=A$; we evaluate the general version with $B \neq A$ in the interest of thoroughness.

We perform the average over $G$ using the same strategy as in App.~\ref{app:k=1 avg} for the onefold channel~\eqref{eq:Avg Chan Def}. First, we write the time-evolved operators $A_G(t)$ and $B_G(t)$~\eqref{eq:noisyOp} in terms of the eigenvalues $\bvec{\lambda}$ and eigenvectors $V$ of $G$~\eqref{eq:onefold in basis}. We then integrate over the diagonalizing unitaries, corresponding to the fourfold Haar average over $\Unitary{N}$. The final step is the average over the eigenvalues of $G$, resulting in SFFs.

Although we are presently concerned with the twofold GUE channel, the procedure we outline applies to generic $k$-fold GUE channels. The first step is to compute the $2k$-fold Haar average over the $V$s. The result involves a sum over $(2k)!^2$ permutations, where  combinatoric factors called Weingarten functions~\cite{Samuel1980, Collins_2003, Collins_2006, BnB} encode the ``weight'' of each permutation. Evaluating the $2k$-fold Haar average over $V \in \Unitary{N}$ leads to
\begin{multline}
    \int_{\Unitary{N}} \hspace{-1mm} \thed \mu(V)\, V^{\smash \vpd}_{\smash a_1 b_1} \cdots V^{\smash \vpd}_{\smash a_{2k} b_{2k}} \,  V^{\smash *}_{\smash \alpha_1 \beta_1} \cdots V^{\smash *}_{\smash \alpha_{2k} \beta_{2k}} = \\
    \hspace{-1.2mm} \sum_{\pi, \tau \in S_{2k}} \hspace{-0.6mm} \Wg(N, \tau^{-1} \pi ) \, \delta_{a_1 \cdots a_{2k}}^{\pi(\alpha_1) \cdots \pi(\alpha_{2k})} \delta_{b_1 \cdots b_{2k}}^{\tau(\beta_1) \cdots \tau(\beta_{2k})} \,, ~\label{eq:2k-fold Haar}
\end{multline}
where $V^{*}$ is the complex conjugate of $V$ (not to be confused with the adjoint $V^\dagger$) and $\pi$ and $\tau$ denote independent permutations of $2k$ objects (the group $S_{2k}$), corresponding to all ways of independently pairing the two indices of $V$s with those of $V^*$s~\cite{BnB, Collins_2003, YoshidaCBD}. Going forward, we omit the explicit dependence on $N$ of the Weingarten function $\Wg$, and refer only to the \emph{cycle structure} of the contractions of $V$s with $V^*$s~\eqref{eq:2k-fold Haar}. For example, the contraction that pairs each $a_j$ and $b_j$ with its corresponding $\alpha_j$ and $\beta_j$, respectively, results in $2k$ cycles of length one---the Weingarten function for this contraction is denoted $\Wg(1^{2k})$. The sum of all cycle lengths is always $2k$. 

For $k>1$, the resulting expressions are quite cumbersome and not particularly enlightening (although we compute an exact expression for $k=2$ below, owing to its relation to the variance). Even enumerating the number of operators on $\Hilbert^{\otimes k}$ as a function of $k$ is challenging. For this reason, the Haar averages described above essentially must be worked out on a case-by-case basis using any computer algebra system. For the reader interested in such calculations, we have included a \texttt{Mathematica} notebook with functions for enumerating all contractions and assigning the appropriate Weingarten functions.

Because the unitaries $V$~\eqref{eq:diagonalizing unitary VG} are contracted with the operators $A_m$ (for $1 \leq m \leq k$) and the eigenvalues $\ee^{\pm \ii t \lambda_n}$, the Haar average~\eqref{eq:2k-fold Haar} results in turn imposes a contraction on the indices of $A_m$ and $\ee^{\pm \ii t \lambda_n}$. It so happens that the Haar average does not mix between these two sets: the components of $A$ and the overall operator on $\Hilbert^{\otimes k}$ may be contracted, and the eigenvalues $\ee^{\pm \ii t \lambda_n}$ may also be independently contracted. The former results in various combinations of the $A_m$ operators and their traces---along with permutation operators on $\Hilbert^{\otimes k}$ that send the state $\ket{a_1, \dots, a_k}$ to some permutation $\ket{\pi ( a_1, \dots, a_k)}$. For $k=2$, there is only the SWAP operator $\mathcal{S} = \sum \hspace{0mm}_{a,b} \ket{ab}\bra{ba}$. The latter contractions result in expressions of the form
\begin{equation}
    \label{eq:2k-fold eig sums}
    \sum\limits_{\substack{m_1, \dots, m_k \\ n_1 , \dots , n_k}} \ee^{\ii t \left( \lambda_{m_1} + \cdots + \lambda_{m_k} - \lambda_{n_1} - \cdots - \lambda_{n_k} \right)}\,  \delta^{\dots}_{\dots} ~ , ~~
\end{equation}
where $ \delta^{\dots}_{\dots}$ denotes arbitrary contractions amongst the indices $m_i$, $n_j$. The result of the contractions is an SFF of the form
\begin{equation}
   \mathcal{R}_{2p,\bvec{q}} (t) = \int \thed \bvec{\lambda} \,  \rho (\bvec{\lambda})  Z(t)^p  Z^* (t)^{p-q} \prod_{i=1}^{n} Z^*(q_i t) \, , ~
    \label{eq:gen SFF}
\end{equation}
where again, $1 \leq p \leq k$, $\bvec{q} = \{q_1, \dots, q_n \}$, $q = \sum_i q_i  < p$, and $Z^*(t)$  is the complex conjugate of $Z(t)$~\eqref{eq:Z(t) def}. Essentially, Kronecker deltas between indices $m_i$ and $n_j$~\eqref{eq:2k-fold eig sums} result in $p<k$, while Kronecker deltas between indices $m_i$ and $m_j$ (or between $n_i$ and $n_j$) result in $q>0$. When $q=0$, $\mathcal{R}_{2p,0}(t)=\mathcal{R}_{2p}(t)$ is the $2p$-point SFF, the Fourier transform of the $2p$-point eigenvalue density function~\cite{YoshidaSFF}. When $q = \sum_i q_i >0$, the result is a $(2p-q)$-point SFF~\cite{YoshidaSFF}---e.g., $\mathcal{R}_{4,1} (t)$~\eqref{eq:SFF3 main} is a three-point SFF. Note that all SFFs~\eqref{eq:gen SFF} are even functions of $t$, meaning that contractions between the $m$ indices or the $n$ indices both result in the same SFF $\mathcal{R}_{2p,\bvec{q}}(t)$~\eqref{eq:gen SFF}. We discuss the evaluation of such higher-point SFFs in Sec.~\ref{subsec:SFF gen}.

However, we comment that some care must be taken in the case $2k > N$~\cite{BnB, Collins_2003, Collins_2006, CDLC1, U1FRUC}. Na\"ively, the generic $2k$-fold Haar average~\eqref{eq:2k-fold Haar} is \emph{only} defined when $2k < N$ (i.e., the number of copies of $V$ and its conjugate are no greater than $N$). However, Ref.~\citenum{Collins_2006} proved that the Haar averages computed for $2k < N$~\cite{BnB, Collins_2003} \emph{also} apply even when $N > 2k$. Most importantly, although the Weingarten functions for the $2k$-fold channel~\eqref{eq:twofold general coefficients} contain denominators with factors $(N^2-j^2)$ for $0 \leq j < 2k$, there are never divergences in the final result~\cite{Collins_2006}. The resolution comes from the SFFs that appear in the $k$-fold GUE channel, which generally results in the $2k$-point and lower SFFs~\eqref{eq:gen SFF}. However, the $p$-point SFF is \emph{only} defined for $p \leq N$; for $p>N$, it reduces to lower-point SFFs multiplied by polynomial functions of $N$. These additional factors of $N$ that come from identifying valid SFFs for $2k>N$ precisely cancel the singular denominators that appear in the Weingarten functions~\eqref{eq:twofold general coefficients}. This occurs, e.g., in the case of qubits, which have a well defined variance---corresponding to the fourfold Haar average---despite the presence of factors of $(N^2-4)$ in the denominators of the Weingarten functions that appear in the twofold GUE channel~\eqref{eq:twofold general coefficients}, even though $2k>N$.

We now return to the particular case of $k=2$ and the twofold channel~\eqref{eq:Var Chan form}. For convenience, we group terms according to their operator content on $\Hilbert^{\otimes 2}$, finding that
\begin{align}
    &\varphi^{(2)}(A,B;t) = C_{A,B} (t) \, A \otimes B + C_{B,A} (t) \, B \otimes A \notag \\
    &+C_{1,A} (t) \left[ \tr (B) \ident \otimes A + \tr (A) B \otimes \ident  + \mathcal{S} \left( \ident \otimes BA + A B \otimes \ident \right) \right] \notag \\
    &+C_{A,1} (t) \left[ \tr (B)  A \otimes 1 + \tr (A) \ident \otimes B + \mathcal{S} \left( \ident \otimes AB + BA \otimes \ident \right) \right] \notag \\
    &+ C_{\tr (A) \tr(B)} (t) \left[ \tr(A) \tr(B)  + \tr(A B) \mathcal{S} \right] \ident^{\otimes 2} \notag \\
    &+ C_{\tr (A B)} (t) \left[ \tr(A B)  + \tr(A) \tr (B) \mathcal{S} \right] \ident^{\otimes 2} \notag \\
    &+ C_{1,AB} (t) \left[ \ident \otimes \left( AB + BA\right) + \left( AB + BA \right) \otimes \ident \right. \notag \\
    &~~~+\left. \tr(B) \mathcal{S} \left( \ident \otimes A + A \otimes \ident \right)  + \tr(A) \mathcal{S} \left( \ident \otimes B + B \otimes \ident \right) \right] \notag \\
    &+ C_{\mathcal{S}} (t) \, \mathcal{S} \left( A \otimes B + B \otimes A \right) \label{eq:twofold general} \, , ~~
\end{align}
where the coefficients $C(t)$ also depend on $N$, and are given by
\begin{widetext}
\begin{align}
    C_{A,B} (t) &= 
    N \Wg (3) 
    - 4 \Wg (1^4) \mathcal{R}_2 (t) + \Wg (2^2) \mathcal{R}_2 (2t) + 2 \Wg (1^2 ,2 ) \mathcal{R}_{4,1}(t) + \Wg (1^4) \mathcal{R}_4(t) \notag \\
    C_{B,A} (t) &= 
     - N \Wg (3) 
     -4 \Wg (2^2) \mathcal{R}_2 (t) 
     + \Wg (1^4) \mathcal{R}_2 (2t) + 2 \Wg (1^2 ,2 ) \mathcal{R}_{4,1}(t) + \Wg (2^2) \mathcal{R}_4(t) \notag \\
    C_{1,A} (t) &= 
    2 \left[ \Wg(1^2,2)  + N \Wg(1,3) \right] \mathcal{R}_2 (t) + \Wg(1^2, 2) \mathcal{R}_2 (2t) 
    + 2 \Wg (1,3) \mathcal{R}_{4,1} (t) + \Wg (4) \mathcal{R}_4 (t) \notag \\
    C_{A,1}(t) &= 
    N \Wg (1,2) + 
    \left[ N \Wg (1^4) - \Wg (1^2,2) \right] \mathcal{R}_2 (t) + \Wg (4) \mathcal{R}_2 (2t)  + 2 \Wg (1, 3) \mathcal{R}_{4,1} (t) + \Wg (1^2, 2) \mathcal{R}_4 (t)  \notag \\
    C_{\tr (A)\tr(B)} (t) &= 
    N \Wg (1^3) 
    - 2 \left[ \Wg (1^4) + \Wg (2^2) \right] \mathcal{R}_2 (t) + \Wg (2^2) \mathcal{R}_2 (2t) + 2 \Wg (4) \mathcal{R}_{4,1} (t) + \Wg (2^2) \mathcal{R}_4(t) \notag \\
    C_{\tr(AB)} (t) &= 
    N \Wg (1,2) 
    - 4 \Wg(1^2,2) \mathcal{R}_2(t)  + \Wg(4) \mathcal{R}_2 (2t) + 2 \Wg(2^2) \mathcal{R}_{4,1}(t) + \Wg(4)  \mathcal{R}_4 (t) \notag \\
    C_{1,AB} (t) &= 
    - \Wg(1,2) 
    - \left[ \Wg (1^4)  + 3 \Wg (1,3) - \Wg (2^2) \right] \mathcal{R}_2 (t)  + \Wg (1,3) \mathcal{R}_{2}(2t) \notag \\
    &\quad ~~~+ \left[ \Wg (1^2,2) + \Wg (4) \right] \mathcal{R}_{4,1}(t) + \Wg (1,3) \mathcal{R}_{4}(t) \notag \\
    C_{\mathcal{S}} (t) &= 
    -4 \Wg (1^2,2) \mathcal{R}_2 (t) + \Wg (1^2,2) \mathcal{R}_2 (2t) + \left[ \Wg (1^4) + \Wg(2^2) \right] \mathcal{R}_{4,1}(t) + \Wg(1^2,2) \mathcal{R}_4 (t) \, . ~~
    \label{eq:twofold general coefficients}
\end{align}
\end{widetext}

\section{Spectral form factors for the GUE}
\label{app:RMT and SFFDetails}

Evaluating the $k$-fold GUE channel $\varphi^{(k)}(A,t)$~\eqref{eq:nfold channel} requires an analytic expression for the $2k$-point SFF, as well as lower-point SFFs. These SFFs also have applications to quantum chaos and even black holes~\cite{BohigasChaos, RMT_SFF, ShenkerRMT, YoshidaSFF, YoshidaCBD, CDLC1, U1FRUC, ConstrainedRUC, MIPT, AmitQuantErgo, delCampoGUESFF, Okuyama_2019}. The SFFs depend only on the eigenvalue density function $\rho (\bvec{\lambda})$~\eqref{eq:spectral pdf}, which for Gaussian ensembles~\cite{mehta2004random, Livan_2018, eynard2018random, YoshidaCBD, TaoRMT, Brezin:2016eax} is given by
\begin{equation*}
\tag{\ref{eq:GaussianSpectralDensity}}
    \rho(\bvec{\lambda}) \equiv \frac{1}{\Omega} \prod_{i<j} \left| \lambda_i-\lambda_j \right|^2 \exp\left(-\frac{\beta}{2} \sum_{i=1}^N \lambda_i^2\right) \, ,~~
\end{equation*}
where we take $\beta=2$ for the GUE. The standard tactic for evaluating integrals over eigenvalue distributions is known as the ``orthogonal-polynomial method''~\cite{mehta2004random}.

\subsection{Orthogonal Polynomials}
\label{app:ortho poly}

Evaluating the standard, two-point SFF $\mathcal{R}_2(t)$~\eqref{eq:SFF main} requires integration of $\cos[t(\lambda_1-\lambda_2)]$ over the GUE eigenvalues $\lambda_1 \neq \lambda_2$. Higher-point SFFs require similar integrals over nonrepeating eigenvalues $\{\lambda_1,\dots,\lambda_m\}$; these integrals can be evaluated using the \emph{marginal} eigenvalue PDF
\begin{equation}
    \rho^{(m)} \left( \lambda_1, \dots, \lambda_m \right) \coloneq \hspace{-0.4mm} \int \hspace{-1mm} \thed \lambda_{m+1} \dots \thed \lambda_N \, \rho (\bvec{\lambda}) \, ,~
    \label{eq:marginal eig pdf general}
\end{equation}
where $\rho (\bvec{\lambda})$~\eqref{eq:spectral pdf} is the full eigenvalue density function. We expect these integrals to be tractable for generic RMT ensembles whose eigenvalue distributions $\rho (\bvec{\lambda})$ are known (e.g., due to rotation invariance), such as the GUE  \eqref{eq:GaussianSpectralDensity}. 

Integrals involving the PDF~\eqref{eq:GaussianSpectralDensity} and its marginals~\eqref{eq:marginal eig pdf general} are simplified substantially 
with the use of a complete, orthonormal basis of (polynomial) functions. 
In particular, the invariance of the Vandermonde determinant $\smash{\Delta(\bvec{\lambda})=\prod_{i<j} \left( \lambda_i - \lambda_j \right) = \det [\mathcal{V}(\bvec{\lambda})]}$ under elementary row operations on the Vandermonde matrix (which satisfies $\mathcal{V}_{ij}=\lambda_i^{j}$ for $0 \leq i,j < N$, with $\lambda_0=1$) \emph{guarantees} that $\smash{\abs{\Delta(\bvec{\lambda})}^2}$ can be conveniently expressed in the form of a finite sum of inner products~\eqref{eq:GUE kernel}, using any basis set of polynomials. The measure $\smash{\ee^{-\lambda^2}}$~\eqref{eq:GaussianSpectralDensity} associated with each eigenvalue $\lambda$ suggests that a pragmatic choice of basis corresponds to the (physicist's) Hermite polynomials $\{H_n(\lambda)\}$, which form a complete orthogonal basis for the Hilbert space $L^2(\Reals)$, i.e.,
\begin{align}\label{eq:Hermite inner}
    \int_{\Reals} \thed x \, \ee^{-x^2} \, H_m (x) H_n(x) = \pi^{1/2} 2^n n! \, \kron{m,n} \, .
\end{align}
For convenience, we use harmonic oscillator eigenstates, whose $n$th \emph{normalized} basis function is given by
\begin{align}\tag{\ref{eq:SHO functions}}
    \phi_n (x) = \inprod{x}{n} = \left(\sqrt{\pi}2^n n!\right)^{-1/2}  \ee^{-x^2/2} H_n(x) \, .
\end{align}

Crucially, writing the marginal PDF $\smash{\rho^{(m)}(\bvec{\lambda})}$~\eqref{eq:marginal eig pdf general} in the basis of Hermite polynomials $H_n(x)$~\eqref{eq:Hermite inner} is particularly convenient. For example, for $m=2$ we have
\begin{align}
\label{appeq:2pointGUE}
    \hspace{-2.25mm} \rho^{(2)} (\lambda_1,\lambda_2) &= \frac{(N-2)!}{N!} \det
    \begin{pmatrix}
        K_N^{i,i} &  K_N^{i,j}\\
         K_N^{j,i} &  K_N^{j,j}
    \end{pmatrix}  , \hspace{-0.5mm}
\end{align}
where the kernel function has a simple representation in terms of the basis functions $\phi_n$~\eqref{eq:SHO functions}, 
\begin{align*}
\tag{\ref{eq:GUE kernel}}
    K_N^{i,j} = K_N(\lambda_i,\lambda_j) \coloneq \sum\limits_{n=0}^{N-1} \phi_{n}(\lambda_i) \phi_{n}(\lambda_j) \, ,~~
\end{align*}
and expressing the $m$-point eigenvalue PDF $\rho^{(m)}(\bvec{\lambda})$~\eqref{eq:marginal eig pdf general} as a determinant~\eqref{appeq:2pointGUE} of the GUE kernel functions $K_N$~\eqref{eq:GUE kernel} leads to a relatively straightforward analytical expression for the SFF. For the two-point SFF~\eqref{eq:SFF main}, e.g., we have
\begin{align}
\label{appeq:SFF2 kappa}
    \mathcal{R}_2(t) = N + N (N-1) \kappa_2 (t) \, ,~~
\end{align}
where we have introduced the function
\begin{align*}
    \kappa_2 (t) \coloneq \int \hspace{-0.7mm} \thed \lambda_1 \thed \lambda_2 \,\rho^{(2)} (\lambda_1, \lambda_2)  \ \ee^{\ii (\lambda_1-\lambda_2) t}  \, ,~
    \tag{\ref{eq:expIntegral}}
\end{align*}
as an integral over $\rho^{(2)}(\lambda_1,\lambda_2)$, which we express in terms of the kernel functions $K_N(\lambda_i,\lambda_j)$~\eqref{appeq:2pointGUE}. Using the harmonic-oscillator eigenbasis for $K_N(\lambda_i,\lambda_j)$~\eqref{eq:GUE kernel}, we arrive at an expression for $\kappa_2 (t)$~\eqref{eq:expIntegral} in terms of integrals of the form
\begin{align}
\label{appeq:SFF matel}
    \int \hspace{-0.4mm} \thed \lambda \, \ee^{\pm \ii \lambda t} \, \phi_m (\lambda) \phi_n (\lambda) = \matel*{m}{\ee^{\pm \ii x t}}{n} \, ,~
\end{align}
where $x$ is the position operator for the harmonic oscillator with Hamiltonian $H=(p^2+x^2)/2$, where $\phi_n(x)$~\eqref{eq:SHO functions} is an eigenstate of this Hamiltonian with eigenvalue $n+1/2$. These methods also extend to higher-point SFFs (see Sec.~\ref{sec:SFF}).

\subsection{Harmonic oscillator matrix elements}
\label{app:SHO matel}

Having reduced generic SFFs to harmonic-oscillator matrix elements of the form $\matel*{m}{\ee^{\pm \ii t x}}{n}$~\eqref{appeq:SFF matel}, we now derive an explicit analytical form for these quantities in terms of generalized Laguerre polynomials. Similar treatments appear in  Refs.~\citenum{schwinger2001quantum} and \citenum{Okuyama_2019}; a proof is given below for convenience. 

Evaluating such matrix elements is most straightforward in the language of the bosonic raising and lowering operators $a$ and $a^\dagger$, respectively, where the lowering operator $a$ acts as
\begin{equation}
\label{eq:lowering operator aciton}
    a^k \ket{m} = \sqrt{\frac{m!}{(m-k)!}} \ket{m-k} \, ,~~
\end{equation}\\
for $k,m \in \Nats$ with $m \geq k$, where $[a,a^\dagger]= \ident$. In terms of this annihilation operator $a$ and its conjugate, the position and momentum operators can be written
\begin{equation}
    \label{eq:x and p from a}
    x = \frac{a+a^\dagger}{\sqrt{2}} ~,~~ p = \frac{a-a^\dagger}{\ii \sqrt{2}} \, ,~~
\end{equation}
so that $[x,p]=\ii \ident$, as expected. A relevant relation is
\begin{equation}
\label{eq:useful exp x relation}
    \ee^{\alpha x} = \ee^{\alpha (a + a^\dagger)/\sqrt{2}} = \ee^{\alpha^2/4} \ee^{\alpha a^\dagger / \sqrt{2}} \ee^{\alpha a / \sqrt{2}}
\end{equation}
and combining this relation (which follows straightfowardly from a Baker-Campbell-Hausdorff expansion) with the action~\eqref{eq:lowering operator aciton} gives for the relevant matrix element~\eqref{eq:expMatEl},
\begin{align}
    \matel*{m}{\ee^{\alpha x}}{n} &= \matel*{m}{\ee^{\alpha^2/4} \ee^{\alpha a^\dagger / \sqrt{2}} \ee^{\alpha a / \sqrt{2}}}{n} \notag \\
    &= \ee^{\frac{\alpha^2}{4}} \sum\limits_{j=0}^{m} \sum\limits_{k=0}^{n}  \frac{1}{j!k!} \left( \frac{\alpha^2}{2}\right)^{\frac{j+k}{2}} \matel*{m}{(a^\dagger )^j a^k}{n} \notag \\
    &= \ee^{\frac{\alpha^2}{4}} \sum\limits_{j,k=0}^{m,n} \frac{\alpha^{j+k}}{j! k!} \sqrt{ \frac{2^{-j-k} \, m! \, n!}{(m-j)!(n-k)!}} \inprod{m-j}{n-k} \, , \notag \\
\intertext{and for $m \geq n$, we have that}
    &= \ee^{\frac{\alpha^2}{4}} \sum\limits_{k=0}^{n} \frac{\alpha^{2k+m-n}}{(k+m-n)! \, k!} 2^{-k} \sqrt{ \frac{2^{n-m} \, m! \, n!}{(n-k)!^2}}  \, ,~\notag \\
    &= \ee^{\frac{\alpha^2}{4}} \sum\limits_{k=0}^{n} \frac{\alpha^{2k+m-n}}{(k+m-n)! \, k!} 2^{-k} \sqrt{ \frac{2^{n-m} \, m! \, n!}{(n-k)!^2}}  \, ,~\notag \\
    &= \ee^{\frac{\alpha^2}{4}} \left( \frac{\alpha}{\sqrt{2}} \right)^{m-n} \sqrt{\frac{n!}{m!}} \sum\limits_{k=0}^{n} \frac{1}{k!} \left( \frac{\alpha^2}{2} \right)^k \begin{pmatrix} m \\ n-k \end{pmatrix} \notag \\
    &= \ee^{\frac{\alpha^2}{4}} \sqrt{\frac{n!}{m!}} \left( \frac{\alpha}{\sqrt{2}} \right)^{m-n} L^{m-n}_n \left( -\frac{\alpha^2}{2} \right) \, ,~ \label{eq:matel Laguerre final m>n}
\end{align}
where $L^{m-n}_n (x)$ is the \emph{generalized} Laguerre polynomial; among other relations, these polynomials satisfy the Rodrigues formula
\begin{align}
    L_n^{(a)}(x) &\coloneq \frac{x^a \, \ee^{x}}{n!} \frac{\thed^n}{\thed x^n} \left( x^{n+a} \ee^{-x} \right) \notag \\
    &\; = \frac{x^{-a}}{n!} \left( \frac{\thed}{\thed x} - 1 \right)^n x^{n+a} \, ,~\label{eq:Laguerre Rodrigues}
\end{align}
for $a \geq 0$; the case $\alpha=0$ realizes the ``simple'' Laguerre polynomials $L_n^{(0)}(x)=L_n(x)$, which form a complete set of orthogonal polynomials with respect to the measure $\ee^{-x}$.

For the case $m \leq n$, we instead have
\begin{align}
    \matel*{m}{\ee^{\alpha x}}{n} &=  \ee^{\frac{\alpha^2}{4}} \sqrt{\frac{m!}{n!}} \left( \frac{\alpha}{\sqrt{2}} \right)^{n-m} L^{n-m}_m \left( -\frac{\alpha^2}{2} \right) \, ,~ \label{eq:matel Laguerre final n>m}
\end{align}
i.e., the roles of $m$ and $n$ are swapped compared to the $m \geq n$ case~\eqref{eq:matel Laguerre final m>n}. One can also derive the $m \leq n$ result~\eqref{eq:matel Laguerre final n>m} from the $m \geq n$ result~\eqref{eq:matel Laguerre final m>n} using the relation  
\begin{equation}
\label{eq:clutchIdentity}
    L^{(-a)}_n (x) = \frac{(n-a)!}{n!} (-x)^a L^a_{n+a}(x) \, ,~~
\end{equation}
for integers $a>0$.

\bibliography{refs}

\end{document}